\documentclass[11pt,nofootinbib,floatfix]{revtex4}
\usepackage{mathrsfs}
\usepackage{graphicx}
\usepackage{amsmath}
\usepackage{amsfonts}
\usepackage{amssymb}
\usepackage{color}
\usepackage[colorlinks=true, linkcolor=blue]{hyperref}

\usepackage{epsfig}
\usepackage{CJK}
\usepackage{eepic}
\usepackage{bbm}
\usepackage{dcolumn}
\usepackage{bm}
\usepackage{ulem}
\usepackage{multirow}

\usepackage{tikz}
\usetikzlibrary{graphs}
\usepackage{minipage-marginpar}
\usepackage{enumitem}
\usepackage{amstext}
\usepackage{braket} 
\usepackage{upgreek}
\usepackage{subcaption}
\usepackage{multirow}
\usepackage{enumitem}
\usepackage{fmtcount}
\usepackage{booktabs}
\usepackage{array}

\newcommand{\omits}[1]{}

\def\bc{\begin{center}}

\def\ec{\end{center}}
\def\be{\begin{eqnarray}}
\def\ee{\end{eqnarray}}

\definecolor{dyellow}{rgb}{1.,0.8,.0}
\definecolor{myblue}{rgb}{.1,.1,.7}
\definecolor{dcyan}{rgb}{.0,.6,.6}
\definecolor{cyan}{rgb}{0.4,1.0,1.0}
\definecolor{dmagenta}{rgb}{0.6,0.0,0.6}
\definecolor{brown}{rgb}{0.6,0.2,0.}
\definecolor{darkblue}{rgb}{.0,.0,0.5}
\definecolor{darkred}{rgb}{0.75,0.0,0.0}
\definecolor{orange}{rgb}{1.,.6,.0}
\definecolor{dorange}{rgb}{0.8,.4,.0}
\definecolor{green}{rgb}{0.0,1.0,0.0}
\definecolor{darkgreen}{rgb}{0.0,0.6,0.0}
\definecolor{purple}{rgb}{.4,.0,.4}
\definecolor{lightgrey}{rgb}{0.7, 0.7, 0.7}
\definecolor{grey}{rgb}{0.4, 0.4, 0.4}

\def\be{\begin{equation}}
\def\ee{\end{equation}}
\def\bea{\begin{eqnarray}}
\def\eea{\end{eqnarray}}

\def\>{\rangle} 
\def\<{\langle} 

\newcommand{\md}{\mathrm d}
\newcommand{\mc}[1]{\mathcal{#1}}

\newcommand{\rsh}{r_{\text{S}}}

\setlist[itemize,1]{label=$\bullet$}   
\setlist[itemize,2]{label=$\circ$}     
\setlist[itemize,3]{label=$\star$}    
\setlist[itemize,4]{label=$\diamond$}

\begin{document}


\title{Acoustic Black Hole in Hayward Spacetime: Shadow, Quasinormal Modes and Analogue Hawking Radiation}

\author{Zhong-Yi Hui$^{1}$} \email{huizhy@mail2.sysu.edu.cn}
\author{Yu-Ye Cheng$^{1}$} \email{chengyy53@mail2.sysu.edu.cn}
\author{Jia-Rui Sun$^{1}$} \email{sunjiarui@mail.sysu.edu.cn}

\affiliation{${}^1$School of Physics and Astronomy, Sun Yat-Sen University, Guangzhou 510275, China}



\begin{abstract}
In this paper, we study an acoustic black hole in Hayward spacetime from the relativistic Gross-Pitaevskii theory. By examining the critical null geodesics, the shadow of the acoustic horizon is sketched. Then the quasinormal mode (QNM) frequencies of the acoustic Hayward black hole are computed numerically using the WKB method, which are shown to be more stable than those of the Hayward black hole, and the variations in the QNM frequencies are shown to correlate with the behavior of the effective potential. Moreover, the WKB method is also employed to calculate the grey-body factor and energy emission rate of the analogue Hawking radiation. It is shown that, as the tuning parameter increases, both the grey-body factor and the energy emission rate are enhanced, which can likewise be attributed to changes in the effective potential. Besides, the radius of acoustic shadow increases with the tuning parameter as well. Our results not only construct an acoustic black hole in regular black hole spacetime, but may also provide potential applications in future observations of astrophysical black holes.
\end{abstract}


\maketitle
\newpage
\tableofcontents

\section{Introduction}
Black holes, among the most important predictions of general relativity, occupy a crucial role both in classical gravity, semi-classical gravity and quantum gravity. Their existence has been confirmed through two landmark observational breakthroughs. First, the LIGO and Virgo Collaboration detected the gravitational wave signal GW 150914 from a binary black hole merger \cite{LIGOScientific:2016vlm}. Subsequently, the Event Horizon Telescope (EHT) obtained direct images of supermassive black hole shadows, initially at the center of galaxy M87* \cite{EventHorizonTelescope:2019dse} and later at the center of our own Milky Way, Sagittarius A* \cite{EventHorizonTelescope:2022wkp}. Together, these observations offer complementary and conclusive evidence for the physical reality of black holes in our universe. Beyond these observational confirmations, black holes are theoretically predicted to exhibit other fundamental properties that are essential to their description. However, quasinormal modes (QNMs) \cite{Press:1971wr, Vishveshwara:1970zz} and Hawking radiation \cite{Hawking:1974rv,Hawking:1975vcx} remain observationally elusive, primarily due to the limited sensitivity of current instruments and the relatively weak energy scales involved. 

In order to observe these phenomena in the laboratory, alternative attempts have been proposed such as probing the Hawking-like radiation in acoustic black hole from analogue gravity~\cite{Barcelo:2005fc}. For example, the sound mode perturbation from a fluid in flat spacetime can be described by a scalar field propagating in an effective curved metric, different velocity profiles of the flow correspond to distinct analogue metrics, thereby enabling the construction of acoustic black holes~\cite{ Unruh:1980cg}. Theoretically, the properties of acoustic black holes, including their horizons, ergospheres and analogue Hawking radiation have been extensively examined~\cite{Visser:1997ux, Vieira:2014rva}. The stability of such systems has also been studied through analysis of QNMs~\cite{Cardoso:2004fi,Berti:2004ju}. Experimentally, the sonic black hole in a Bose-Einstein condensate (BEC) has been realized in~\cite{Lahav:2009wx}, followed by observations of analogue Hawking radiation and its corresponding Hawking temperature~\cite{MunozdeNova:2018fxv,Isoard:2019buh}. Moreover, progresses on simulated Hawking emission have been made in optical platforms~\cite{Steinhauer:2014dra,Drori:2018ivu}. A comprehensive overview of both theoretical insights and experimental advances in this field has been provided in the review~\cite{Barcelo:2005fc}.

On the other hand, analogue gravitational systems not only exist in flat spacetimes, but their typical features can also appear in curved spacetimes. For example, acoustic black hole geometries can be derived from the relativistic Gross-Pitaevskii (GP) theory and Yang-Mills (YM) theory in the real black hole spacetime backgrounds~\cite{Ge:2019our}. Recent works have constructed acoustic black holes within Schwarzschild and charged black hole backgrounds, analyzing various interesting properties including quasi-bound states \cite{Vieira:2023ylz,Vieira:2021xqw}, QNMs, Hawking radiation, and black hole shadows \cite{Guo:2020blq,Ling:2021vgk}. Moreover, the information paradox of acoustic black hole in Schwarzschild spacetime has been investigated using the island formula \cite{Cheng:2025bnj}. The study of acoustic black holes in black hole background spacetimes is of notable physical significance, as real astrophysical black holes likely reside in environments resembling normal fluid, superfluid medium or the cosmic microwave background. The presence of an acoustic horizon can alter dynamical behavior near the gravitational horizon, potentially providing new insights and clues for observational studies of astrophysical black holes.

The discussions of acoustic black holes in curved spacetimes remain confined to singular background spacetimes. It is therefore of interest to extend such studies to regular black hole backgrounds, which refer to a class of black hole solutions that possess event horizons but are free of essential singularities throughout the entire spacetime \cite{Hayward:2005gi,Dymnikova:1992ux,Frolov:2016pav}. These solutions can be constructed from models within nonlinear electrodynamics \cite{Ayon-Beato:1998hmi,Fan:2016hvf} and modified gravity theories \cite{Bueno:2024dgm}. Many important properties of regular black holes have been extensively studied, such as particle geodesics \cite{Abbas:2014oua, Stuchlik:2014qja}, QNMs \cite{Fernando:2012yw,Flachi:2012nv,Lin:2013ofa}, and thermodynamic behaviors \cite{Molina:2021hgx}. Moreover, the microlensing of regular black holes has been investigated via ray-tracing methods \cite{Boos:2025nzc}.

In the present paper, we aim to investigate the near-horizon properties associated with observable quantities of the acoustic black holes within the Hayward black hole background~\cite{Hayward:2005gi}, which is a typical class of regular black holes that attracting many attentions recently. In other words, we will investigate the so-called acoustic Hayward black hole, and our study will focuses on the acoustic shadow, QNMs, grey-body factors, and the energy emission rate of analogue Hawking radiation for this system. The acoustic shadow can be studied through phonon propagation, which is effectively described by null geodesics in the analogue gravity framework. Meanwhile, QNMs and analogue Hawking radiation can be analyzed via scalar fields within the same analogue gravity formalism. Explicitly, by examining the critical null geodesics, the shadow of the acoustic horizon is sketched. Then the QNM frequencies of the acoustic Hayward black hole are computed numerically using the WKB method, confirming the stability of the system, and variations in the QNM frequencies are shown to correlate with the behavior of the effective potential. Moreover, the WKB method is also employed to calculate the grey-body factor and energy emission rate of the analogue Hawking radiation. Our results show that as the tuning parameter increases, the greybody factor, the energy
emission rate, and shadow radius are all significantly increased. In contrast, the absolute value of the QNM decreases accordingly, and the potential function becomes smoother, indicating that the QNM tends to stabilize in the acoustic scenario. By comparison, variations in the Hayward parameter make only a slight influence on the above physical quantities, which demonstrates the uniqueness of acoustic black holes. Our results not only extend the acoustic black hole in regular black hole spacetime, but may also be useful in future experimental tests of astrophysical black holes.

The structure of this paper is organized as follows: In Sec.~\ref{sec:2}, we construct the metric of the acoustic Hayward black hole and analyze its horizon structure. After a brief review of the horizon structure of Hayward black hole, we recall the formulation of analogue gravity within the GP theory and apply it in the Hayward spacetime background. In Sec.~\ref{sec:3}, the acoustic shadow of the acoustic Hayward black hole is examined by analyzing null geodesics. The variation of the shadow radius with different parameters is discussed. Subsequently, in Sec.~\ref{sec:Cov_eq} we analyze the covariant scalar field equation and the effective potential in acoustic Hayward spacetime, laying the groundwork for subsequent calculations. In Sec.~\ref{sec:QNMs}, the WKB method is employed to numerically compute QNMs under various parameters, and the stability of the acoustic Hayward black hole is examined. Then, in Sec.~\ref{sec:aHR} the grey-body factors and energy emission rate in analogue Hawking radiation are calculated, also using the WKB method. Finally, Sec.~\ref{sec:con} presents the conclusions and discussions.
\section{Acoustic Hayward black hole}\label{sec:2}

\subsection{Hayward black hole}

The Hayward black hole is a typical example of regular black holes, which can be constructed via nonlinear electrodynamics, the action for an electrically charged black hole is \cite{Fan:2016hvf}
\begin{equation}\label{eq:action_nled}
    I=\frac{1}{16\pi}\int\md^4 x\sqrt{-g}(R-\mathcal{L}(\mathfrak{F})),
\end{equation}
here we have defined $\mathfrak F \equiv F_{\mu\nu} F^{\mu\nu}$, where $F= \md A$ is the field strength of the electromagnetic field associated with the vector potential $A^\mu$, and  $\mathcal{L}(\mathfrak{F})$ is the electromagnetic Lagrangian density. According to the principle of least action, the gravitational field equations are
\begin{equation}\label{eq:ein_eq}
    G_{\mu\nu}=T_{\mu\nu},
\end{equation}
in which $G_{\mu\nu}=R_{\mu\nu}-\frac{1}{2}Rg_{\mu\nu}$ is the Einstein tensor, and the stress tensor is
\begin{equation}
    T^\nu_\mu=2\mathcal{L}_{\mathfrak{F}}F_{\mu\alpha}F^{\nu\alpha}-\frac{1}{2}\delta^\nu_\mu\mathcal{L},
\end{equation}
where $\mathcal{L}_{\mathfrak{F}}:=\frac{\partial\mathcal{L}}{\partial \mathfrak{F}}$. And the equations of motion  for the electromagnetic field are
\begin{equation}
    \nabla_\mu\left(\mathcal{L}_{\mathcal{F}}F^{\mu\nu}\right)=0
\end{equation}
combined with the Bianchi identities
\begin{equation}\label{eq:bianchi}
    \nabla_\mu(\star F^{\mu\nu})=0,
\end{equation}
the electromagnetic field can be fully determined, where $\star$ denotes the Hodge dual. Therefore, for nonlinear electrodynamics, different electromagnetic Lagrangians $\mathcal{L}(\mathfrak{F})$ lead to distinct spacetimes and electromagnetic fields \cite{Bronnikov:2000vy, Fan:2016hvf}. Especially, the Hayward black hole can be constructed as a purely magnetically charged black hole, whose Lagrangian density is \cite{Fan:2016hvf}
\begin{equation}
    \mc{L}=\frac{6}{L^2}\frac{(2L^2\mathfrak{F})^{\frac{3}{2}}}{[1+(2L^2\mathfrak{F})^{\frac{3}{4}}]^2},
\end{equation}
where the Hayward parameter $L>0$ is a length-scale parameter \cite{Frolov:2016pav} and is related to the magnetic charge $q_m$ of the Hayward black hole via $q_m=\frac{1}{2}\sqrt[3]{r_s^2L}$, with $r_s=2M$ denoting the Schwarzschild radius \cite{Fan:2016hvf}. As a result, the Hayward metric is given by
\begin{equation}\label{eq:hayward_spacetime}
    \md s^2=-f(r)\md t^2+f^{-1}(r)\md r^2+r^2\left(\md\theta^2+\sin^2\theta\md\phi^2\right),
\end{equation}
with the lapse function given by \cite{Hayward:2005gi}
\begin{equation}\label{eq:hayward_lapse}
    f(r)=1-\frac{r_s r^2}{r^3+r_s L^2}:=1-g(r).
\end{equation}
For the subsequent discussion, we define $g(r):=\frac{r_s r^2}{r^3+r_s L^2}$. This Hayward black hole indeed avoids the intrinsic singularity at its center, since as $r\rightarrow0$, the lapse function behaves as $f(r)\rightarrow1-\frac{r^2}{L^2}$, exhibiting the form of de Sitter spacetime core with finite curvature. Meanwhile, at spatial infinity, as $r\rightarrow\infty$, $f(r)\rightarrow 1-\frac{r_s}{r}$, matching the asymptotic form of the Schwarzschild spacetime.

To analyze the horizon structure of the Hayward black hole, we rewrite the lapse function $f(r)$ in polynomial form: \cite{Molina:2021hgx}
\begin{equation}
    f(r)=1-\frac{r_sr^2}{r^3+r_sL^2}=\frac{P_3(r;r_s,L)}{r^3+r_sL^2},
\end{equation}
where 
\begin{equation}\label{eq:p3r}
    P_3(r;r_s,L):=r^3-r_sr^2+r_sL^2,
\end{equation}
which is an ordinary cubic polynomial. To find its roots, a standard technique is to shift the variable $r$ via $r=x+\frac{r_s}{3}$, which transforms $P_3$ into a polynomial in $x$ with missing terms: 
\begin{equation}\label{eq:p3x}
    P_3(x;r_s,L)\equiv x^3-\frac{g_2}{4}x-\frac{g_3}{4},
\end{equation}
where the coefficients $g_2$ and $g_3$ are given by
\begin{equation}
    g_2=\frac{4r_s^2}{3}>0,\quad g_3=\frac{8r_s^3}{27}\left(1-\frac{L^2}{L_0^2}\right)=\frac{1}{\sqrt{27}}g_2^{\frac{3}{2}}\left(1-\frac{L^2}{L_0^2}\right),
\end{equation}
with the constant $L_0$ defined as
\begin{equation}
    L_0\equiv\sqrt{\frac{2}{27}}r_s\approx 0.272r_s.
\end{equation}
The cubic discriminant of Eq.~\eqref{eq:p3x} is given by \cite{lang2012algebra}
\begin{equation}\label{eq:discriminant}
    \Delta_c=-4\left(-\frac{g_2}{4}\right)^3-27\left(-\frac{g_3}{4}\right)^2
    =\frac{g_2^3}{16}\left[1-\left(1-\frac{L^2}{L_0^2}\right)^2\right],
\end{equation}
and the sign of the discriminant $\Delta_c$ determines the nature of the roots as follows: \cite{Molina:2021hgx}
\begin{itemize}
    \item For $\Delta_c>0\ (L<\sqrt{2}L_0)$, there are three distinct real roots: two postive and one negative;
    \item For $\Delta_c=0\ (L=0, \sqrt{2}L_0)$, there are one positive double root and one distinct negative root;
    \item For $\Delta_c<0\ (L>\sqrt{2}L_0)$, there are one negative root and a complex conjugate pair.
\end{itemize}
Therefore, Hayward black hole possesses an event horizon only if $\Delta_c\geq0$, that is $0\leq L\leq\sqrt{2}L_0$. The horizon radius is determined by solving the equation
\begin{equation}\label{eq:4x3}
    4x^3-g_2x-g_3=0.
\end{equation} 
With the substitution $x=\sqrt\frac{g_2}{3}\cos\alpha$, the equation simplifies to
\begin{equation}\label{eq:4cos3}
    4\cos^3\alpha-3\cos\alpha-\left(1-\frac{L^2}{L_0^2}\right)=0.
\end{equation}
Using the trigonometric identity 
\begin{equation}\label{eq:tri_identity}
    4\cos^3\alpha-3\cos\alpha-\cos 3\alpha=0,
\end{equation}
we further obtain
\begin{equation}\label{eq:cos3a}
    \cos3\alpha=1-\frac{L^2}{L_0^2}.
\end{equation}
For $\Delta_c\geq0$, we have $-1\leq1-\frac{L^2}{L_0^2}\leq1$, which ensures that $\alpha$ is real. The general solution can be written as
\begin{equation}\label{eq:3alparc}
    3\alpha=\arccos\left(1-\frac{L^2}{L_0^2}\right)+2n\pi,\quad n=0,1,2.
\end{equation}
The outermost root, corresponding to the event horizon, is obtained for $n=0$, giving
\begin{equation}\label{eq:sol_x}
    x=\sqrt{\frac{g_2}{3}}\cos\alpha,\quad\alpha=\frac{1}{3}\arccos\left(1-\frac{L^2}{L_0^2}\right).
\end{equation}
Finally, transforming back to the original variable, the event horizon radius of Hayward black hole is
\begin{equation}\label{eq:hayward_horizon}
    r_{\text H}=\frac{r_s}{3}\left(2\cos\alpha+1\right),\quad\alpha=\frac{1}{3}\arccos\left(1-\frac{L^2}{L_0^2}\right).
\end{equation}

\subsection{Analogue gravity from the GP theory}
Returning to the construction of the acoustic black hole from relativistic GP theory \cite{Ge:2019our}, we consider fluctuations of a complex scalar field $\varphi$. The action of GP theory is given by \cite{Gross:1961kqh, pitaevskii1961vortex}
\begin{equation}
    S=\int\md^4 x\sqrt{-g}\left(|\partial_\mu\varphi|^2+m^2|\varphi|^2-\frac{b}{2}|\varphi|^4\right),
\end{equation}
where $b$ is a constant, and $m^2$ is a temperature dependent parameter that governs the behavior of the system near the critical temperature $T_c$, scaling as $m^2\sim (T-T_c)$. In a static background spacetime with metric
\begin{equation}\label{eq:bgspacetime}
    \md s^2_{\text{bg}}=g_{tt}\md t^2+g_{xx}\md x^2+g_{yy}\md y^2+g_{zz}\md z^2,
\end{equation}
the equation of motion for $\varphi$ follows as
\begin{equation}\label{eq:kgeq}
    \square\varphi+m^2\varphi-b|\varphi|^2\varphi=0,
\end{equation}
where $\square\varphi:=\frac{1}{\sqrt{-g}}\partial_\mu\left(\sqrt{-g}g^{\mu\nu}\partial_\nu\varphi\right)$. Using the Madelung representation $\varphi=\sqrt{\rho(\bm x,t)}e^{i\theta(\bm x,t)}$ the complex field $\varphi$ is separated, leading the equation of motion \eqref{eq:kgeq} decomposing into its real and imaginary parts. The real component gives the Euler equation
\begin{equation}\label{eq:euler}
        \frac{1}{\sqrt{\rho}}\frac{1}{\sqrt{-g}}\partial_\mu\left(\sqrt{-g}g^{\mu\nu}\partial_\nu\sqrt{\rho}\right)-g^{\mu\nu}(\partial_\mu\theta)(\partial_\nu\theta)+m^2-b\rho=0, 
\end{equation}
where the first term corresponds to the quantum potential $\frac{\square\sqrt\rho}{\sqrt\rho}$. In the long-wavelength approximation adopted here, this quantum potential term is neglected \cite{Ge:2019our}. Meanwhile, the imaginary part takes the form
\begin{equation}
    \frac{1}{\sqrt{-g}}\partial_\mu\left(\sqrt{-g}g^{\mu\nu}\sqrt{\rho}\partial_\nu\theta\right)+g^{\mu\nu}\left(\partial_\mu\sqrt{\rho}\right)\left(\partial_\nu\theta\right)=0,
\end{equation}
which simplifies to the equation of continuity
\begin{equation}\label{eq:continius}
    \frac{1}{\sqrt{-g}}\partial_\mu\left(\sqrt{-g}g^{\mu\nu}\rho\partial_\nu\theta\right)=0.
\end{equation}

Consider small fluctuations $(\rho_1,\theta_1)$ around a background configuration $(\rho_0,\theta_0)$, i.e., $\rho=\rho_0+\rho_1$ and $\theta=\theta_0+\theta_1$. Linearizing Eqs.~\eqref{eq:euler} and \eqref{eq:continius} yields, at leading order,
\begin{equation}\label{eq:brho0v}
    b\rho_0=m^2-g^{\mu\nu}(\partial_\mu\theta_0)(\partial_\nu\theta_0)=m^2-v_\mu v^\mu,
\end{equation}
where $v_\mu$ is defined as $v_t:=-\partial_t\theta_0$ and $v_i:=\partial_i\theta_0$. At sub-leading order, we obtain
\begin{equation}\label{eq:covariant_eq_theta}
    \frac{1}{\sqrt{-\mathcal{G}}}\partial_\mu\left(\sqrt{-\mathcal{G}}\mathcal{G}^{\mu\nu}\partial_\nu\theta_1\right)=0,
\end{equation}
which is a relativistic D'Alembertian equation for the phase fluctuation $\theta_1$, i.e., the phonon, propagating in an effective curved spacetime described by the effective metric $\mathcal{G}_{\mu\nu}$, which is
\begin{equation}
    \mathcal{G}_{\mu\nu}=\frac{c_s}{\sqrt{c_s^2-v_\mu v^\mu}}\left[\begin{array}{ccc}
        g_{tt}(c_s^2-v_iv^i) & \vdots & -v_iv_t \\
        \cdots \cdots  & \cdot &\cdots \cdots \cdots\cdots\cdots\cdots \\
        -v_iv_t & \vdots & g_{ii}(c_s^2-v_\mu v^\mu)\delta^{ij}+v_iv_j
        \end{array}\right],
\end{equation}
with $c_s^2:=\frac{1}{2}b\rho_0$. It is evident that the effective metric $\mathcal{G}_{\mu\nu}$ depends on both the background spacetime $\md s^2_{\text{bg}}$ and the four-velocity $v_\mu$ of the fluid. When the four-velocity has only $t$ and $r$ components, namely, $v_t\neq0,v_r\neq0$, and $v_\theta=v_\phi=0$, the effective metric can be further reduced to \cite{Ge:2019our}: 
\begin{equation}
    \mathcal{G}_{\mu\nu}=\frac{c_s}{\sqrt{c_s^2-v_\mu v^\mu}}
    \left[\begin{array}{cccc}
        g_{tt}(c_s^2-v_rv^r) & -v_tv_r & 0 & 0 \\
        -v_tv_r & g_{rr}(c_s^2-v_tv^t) & 0 & 0 \\
        0 & 0 & g_{\theta\theta}(c_s^2-v_\mu v^\mu) & 0 \\
        0 & 0 & 0 & g_{\phi\phi}(c_s^2-v_\mu v^\mu)
    \end{array}\right].
\end{equation} 
To proceed, we diagonalize the effective metric via the coordinate transformation $(\md t-\frac{v_tv_r}{c_s^2-v_rv^r}\md r)\rightarrow\md t$. Imposing the condition $g_{tt}g_{rr}=-1$ for the background spacetime, the effective line element becomes
\begin{equation}\label{eq:eff_line}
    \md s^2=c_s\sqrt{c_s^2-v_\mu v^\mu}\left(\frac{c_s^2-v_rv^r}{c_s^2-v_\mu v^\mu}g_{tt}\md t^2+\frac{c_s^2}{c_s^2-v_r v^r}g_{rr}\md r^2+g_{\theta\theta}\md\theta^2+g_{\phi\phi}\md\phi^2\right).
\end{equation}
The analogue gravitational metric emerging from the relativistic GP theory is thus derived.
\subsection{Acoustic Hayward black hole}
We now embed the effective metric Eq.~\eqref{eq:eff_line} into the Hayward background spacetime described by Eq.~\eqref{eq:hayward_spacetime}. This corresponds to studying the acoustic black hole arising in the fluid governed by GP theory in the vicinity of a Hayward black hole. In other words, we construct the acoustic Hayward black hole. To determine the effective metric in analogue gravity, the four-velocity of the fluid must be specified. In this paper, we consider a fluid that starts at rest at infinity and falls radially freely outside a Hayward black hole. Thus, the radial component $v_r$ corresponds to the escape velocity required to remain stationary at a given radius $r$, and is given by $v_r=\sqrt{g(r)\xi}$, where $\xi>0$ acts as a tuning parameter, and its range for the existence of an acoustic horizon will be discussed later. By rescaling $m^2\rightarrow\frac{m^2}{2c_s^2}$ and $v_\mu v^\mu\rightarrow \frac{v_\mu v^\mu}{2c_s^2}$, the relation $v_\mu v^\mu=m^2-1$ is obtained. At the critical temperature of GP theory, where $m^2$ vanishes, it follows that $v_\mu v^\mu=-1$. Setting $c_s=\frac{1}{\sqrt{3}}$ for convenience and substituting the four-velocity together with the Hayward background spacetime \eqref{eq:hayward_spacetime} into the effective line element \eqref{eq:eff_line},
the line element of acoustic Hayward black hole is derived as
\begin{equation}\label{eq:acoustic_hayward_spacetime}
    \md s^2=-\mc{F}(r)\md t^2+\frac{1}{\mc F(r)}\md r^2+r^2(\md\theta^2+\sin^2\theta\md\phi^2),
\end{equation}
where
\begin{equation}\label{eq:Fr}
    \mc F(r)=\left(1-\frac{r_sr^2}{r^3+r_sL^2}\right)\left[1-\xi\frac{r_sr^2}{r^3+r_sL^2}\left(1-\frac{r_sr^2}{r^3+r_sL^2}\right)\right].
\end{equation}
Using the definition of $g(r)$ in Eq.\eqref{eq:hayward_lapse}, it can be written more compactly as
\begin{equation}\label{eq:mcF}
    \mc F(r)=\big(1-g(r)\big)\big[1-\xi g(r)\big(1-g(r)\big)\big].
\end{equation}
We can see that the acoustic Hayward black hole described by Eq.~\eqref{eq:acoustic_hayward_spacetime} reduces to the original Hayward black hole Eq.~\eqref{eq:hayward_spacetime} in the limit $\xi\rightarrow 0$. Conversely, as  $\xi\rightarrow\infty$, the acoustic Hayward black hole fills the entire spacetime, which will be examined in more detail in the subsequent discussion.

The acoustic horizon is determined by solving $\mc F(r)=0$. Vanishing of the first term in Eq.~\eqref{eq:mcF} gives the horizon of the background Hayward black hole Eq.~\eqref{eq:hayward_horizon}, while setting the second term to zero gives the acoustic horizon of the acoustic Hayward black hole, which satisfies
\begin{equation}
    1-\xi g(r)\big(1-g(r)\big)=0.
\end{equation}
This can be simplified to
\begin{equation}\label{eq:grlam}
    g(r)=\frac{1}{2}\pm\sqrt{\frac{1}{4}-\frac{1}{\xi}},
\end{equation}
and we define the right term $\frac{1}{2}\pm\sqrt{\frac{1}{4}-\frac{1}{\xi}}:=\lambda_\pm$. This result also implies the constraint $\xi\geq4$ for the existence of acoustic black hole. By substituting the expression of $g(r)$ defined in Eq.~\eqref{eq:hayward_lapse} into Eq.~\eqref{eq:grlam}, we obtain the following cubic equation for the acoustic horizon in terms of $r$:
\begin{equation}\label{eq:r3lam}
    r^3-\frac{r_s}{\lambda_{\pm}} r^2+r_sL^2=0,
\end{equation}
which has a similar form to the Hayward horizon equation \eqref{eq:p3r}. Therefore, we can apply the same method by making the substitution $r=x+{\frac{r_s}{3\lambda_\pm}}$ to remove the quadratic term. After this transformation, Eq.~\eqref{eq:r3lam} becomes
\begin{equation}
    x^3-\frac{r_s^2}{3\lambda^2}x+\left(r_sL^2-\frac{2r_s^3}{27\lambda^3_\pm}\right)={x^3-\frac{\tilde{g}_2}{4}x-\frac{\tilde{g}_3}{4}=0},
\end{equation}
where the coefficients $\tilde{g}_2$ and $\tilde{g}_3$ are defined as
\begin{equation}
\tilde{g}_2=\frac{4r_s^2}{3\lambda_\pm^2}>0,\quad \tilde{g}_3=\frac{8r_s^3}{27\lambda_\pm^3}\left(1-\frac{L^2}{(\tilde{L}_\pm)^2}\right)=\frac{1}{27}(\tilde{g}_2)^{\frac{3}{2}}\left(1-\frac{L^2}{(\tilde{L}_\pm)^2}\right),
\end{equation}
and $\tilde{L}_\pm=\sqrt{\frac{2}{27\lambda^3_\pm}}r_s$. Similarly, the cubic discriminant is
\begin{equation}
    \tilde\Delta_c=-4\left(-\frac{\tilde g_2}{4}\right)^3-27\left(-\frac{\tilde g_3}{4}\right)^2=\frac{(\tilde g_2)^3}{16}\left[1-\left(1-\frac{L^2}{(\tilde{L}_\pm)^2}\right)^2\right].
\end{equation}
The acoustic horizon exists only when $0\leq L\leq\sqrt{2}\tilde L_\pm$, which ensures the discriminant $\tilde\Delta_c\geq0$. Following the same procedure as in Eq.~\eqref{eq:4cos3} to Eq.~\eqref{eq:3alparc}, the solutions of $x$ are given by
\begin{equation}
    x=\sqrt{\frac{\tilde g_2}{3}}\cos\tilde\alpha_\pm,\quad\tilde\alpha_\pm=\frac{1}{3}\arccos\left(1-\frac{L^2}{(\tilde L_\pm)^2}\right).
\end{equation}
And the acoustic Hayward horizons are finally obtained as
\begin{equation}
    r_\mp'=\frac{r_s}{3\lambda_\pm}\left(1+2\cos\tilde\alpha_\pm\right),\quad\tilde\alpha_\pm=\frac{1}{3}\arccos\left(1-\frac{L^2}{(\tilde L_\pm)^2}\right).
\end{equation}
Given that $0<\lambda_-\leq\frac{1}{2}\leq\lambda_+<1$, we have $ r_+'\geq  r_-'$, with equality holding  at $\xi =4$, corresponding to the extremal acoustic Hayward black hole. Moreover, considering both the Hayward black hole and the acoustic black hole simultaneously, the conditions $\Delta_c\geq0$ and $\tilde\Delta_c\geq0 $ imply $0\leq L\leq \sqrt{2}L_0$. To compare the relative sizes of $r_{\text H}$ and $ r_\pm'$, note that the relation $-1<1-\frac{L^2}{L_0^2}<1-\frac{L^2}{(\tilde L_\pm)^2}<1$ gives $\cos\alpha<\cos\tilde\alpha_\pm$, which guarantees that the acoustic horizons $ r_\pm'$ always lie outside the Hayward $r_\text H$ horizon as expected. In summary, we obtain that
\begin{equation}
    r_+'>r'_->r_{\text H}.
\end{equation}

\begin{figure}[htbp]
    \begin{minipage}[b]{0.48\columnwidth}
        \centering
        \includegraphics[width=\linewidth]{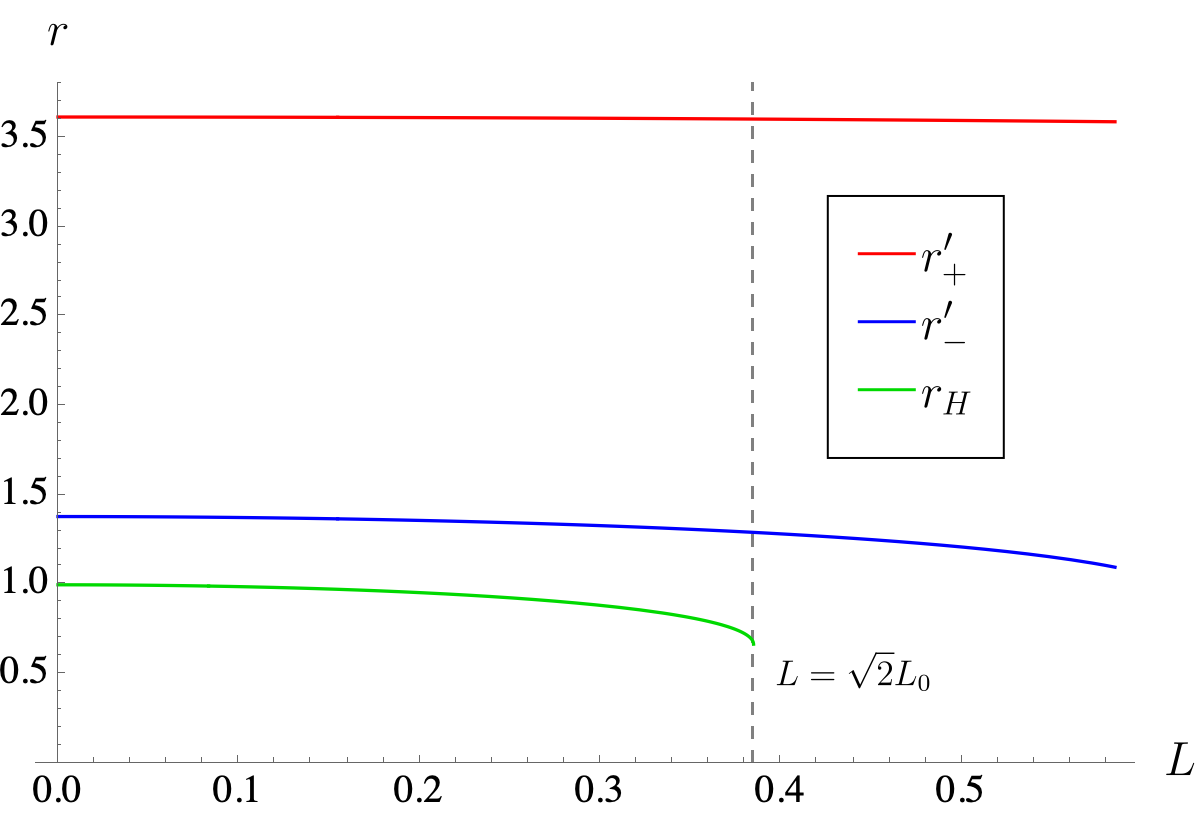}
        \par\vspace{3pt}
    \end{minipage}
    \hfill 
    \begin{minipage}[b]{0.48\columnwidth}
        \centering
        \includegraphics[width=\linewidth]{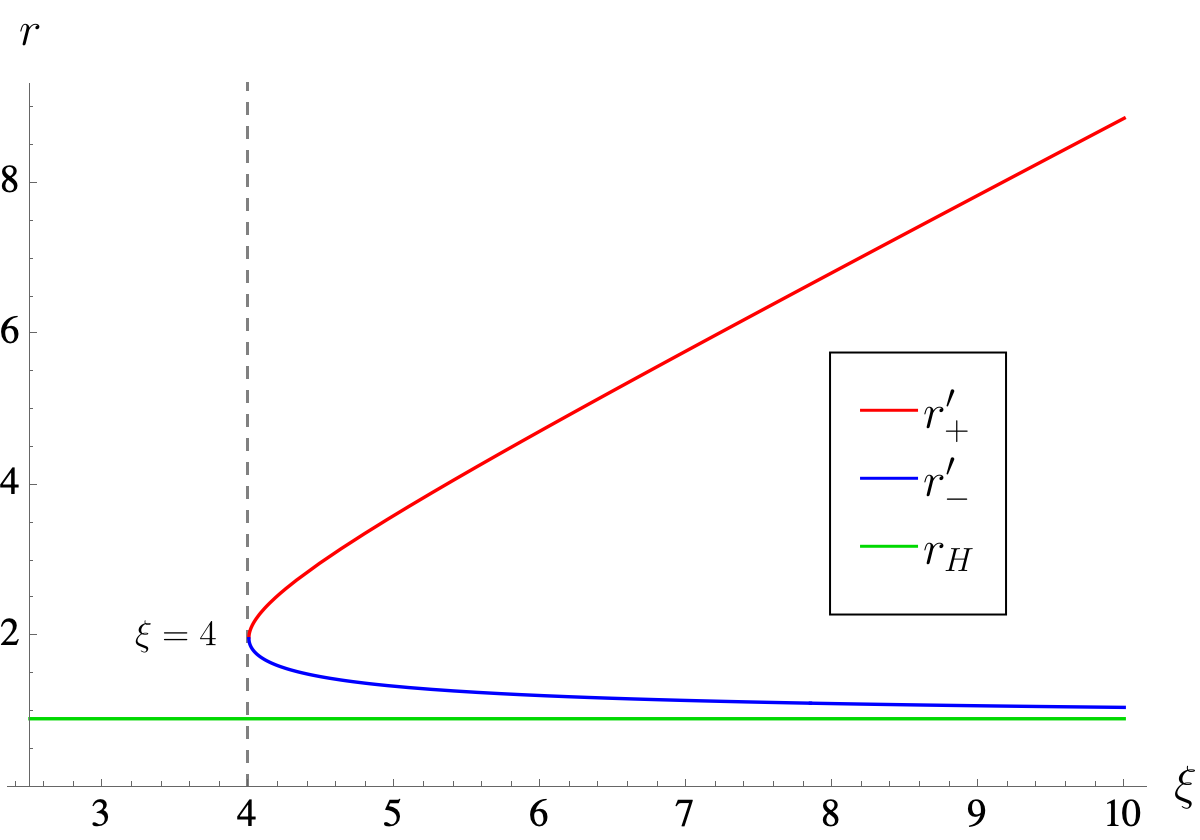}
        \par\vspace{3pt}
    \end{minipage}
    
    \caption{\raggedright Horizon structure of the acoustic Hayward black hole. Left panel: Radii of the outer acoustic horizon $r_+'$, inner acoustic horizon $r_-'$, and Hayward horizon $r_{\text H}$ as functions of $L$, with fixed $\xi=5$. The vertical dashed line indicates the critical value $L=\sqrt{2}L_0$. Right panel: The same horizon radii as functions of $\xi$ for a fixed $L=L_0$. The vertical dashed line marks the critical value $\xi=4$.}
    \label{fig:horizon}
\end{figure}

The Hayward horizon and acoustic horizons are illustrated in Fig.~\ref{fig:horizon}. Without loss of generality, we set $r_s=1$ in the following discussion. In the left panel, with $\xi=5$ fixed, we show how the horizon radii vary with the Hayward parameter $L$. One can see that $r_+'>r'_->r_{\text H}$, and that the Hayward black hole exists for $0\leq L\leq\sqrt{2}L_0$. In the right panel, with $L=L_0$ fixed, we plot the horizon radii as functions of the tuning parameter $\xi$. The acoustic horizons $r_\pm'$ exist for $\xi\geq4$, and the case $\xi=4$ corresponds to the extremal acoustic black hole, as noted earlier. In the limit $\xi\rightarrow\infty$, we have $r'_+\rightarrow\infty$ indicating that sound cannot escape from the entire spacetime, which is consistent with the earlier discussion. In summary, for $\xi\geq 4$ and $0\leq L\leq\sqrt{2}L_0$, the analogue metric embedded in the Hayward spacetime partitions the spacetime into four regions:
\begin{itemize}
    \item $r<r_{\text H}$: the interior of Hayward black hole;
    \item $r_{\text H}<r<r_-'$: a region where light can escape, and sound can also escape but cannot be detected by an observer outside this region;
    \item $r_-'<r<r_+'$: a region where light can escape but sound cannot;
    \item $r>r_+'$: a region where both light and sound can escape.
\end{itemize}
In the subsequent analysis, the acoustic horizon is designated as the outer horizon $r_+'$.
\section{Acoustic black hole shadow}\label{sec:3}
\begin{figure}[ht]
    \centering
    \includegraphics[scale=0.45]{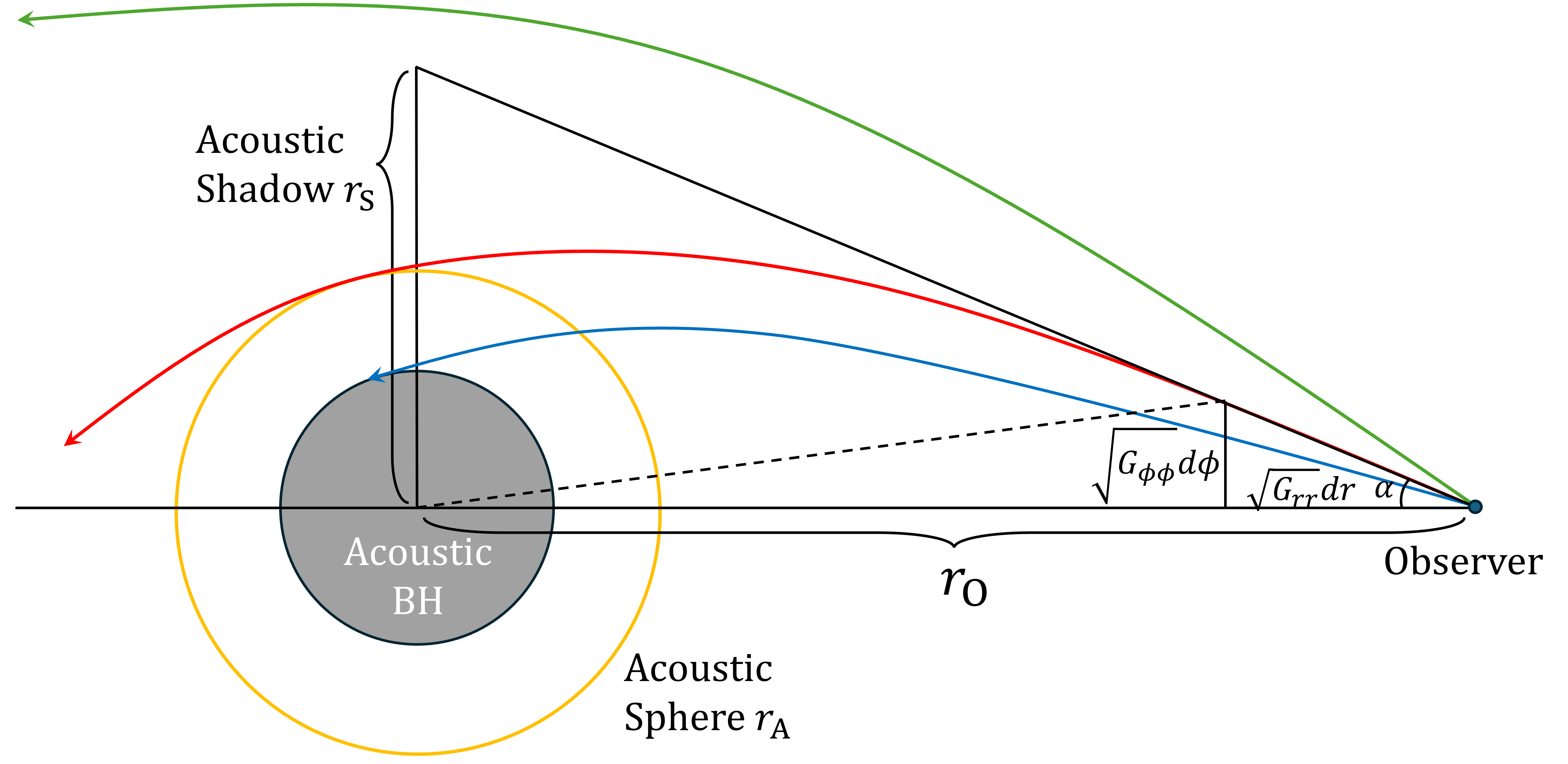}
    \caption{\raggedright Schematic of the acoustic black hole shadow. The central gray circle denotes the acoustic black hole, surrounded by an acoustic sphere (yellow circle) of radius $r_\text A$. Null geodesics emitted from the observer on the right are grouped into three families: blue trajectories captured by the acoustic black hole; green trajectories escape after deflection; and red critical geodesics that determine the acoustic shadow between the captured and escaping paths.}
    \label{fig:bh_shadow}
\end{figure}

We begin by investigating the acoustic shadow of the acoustic Hayward black hole. In analogue gravity, sound rays follow the null geodesics of the acoustic metric \cite{Barcelo:2005fc}. Consequently, an acoustic black hole also exhibits an acoustic shadow for a distant observer. The formation of such a shadow can be understood with the help of Fig.~\ref{fig:bh_shadow}. The grey region represents the acoustic black hole, and the rightmost point corresponds to the observer. Observation of the acoustic black hole by the observer can be described in terms of null geodesics. As illustrated, these null geodesics fall into two categories: those shown in blue and those in green. The blue geodesics are ultimately captured by the acoustic black hole, whereas the green ones, though deflected, manage to escape. The boundary between captured and escaping trajectories is defined by the red geodesics, which play a pivotal role in the formation of the acoustic shadow. In what follows, we shall therefore concentrate our analysis on this particular class of critical null geodesic.

For geodesics, the Lagrangian takes the form
\begin{equation}
    \mathcal{L}(x,\dot{x})=\frac{1}{2}\mathcal{G}_{\mu\nu}\dot{x}^\mu\dot{x}^\nu=\frac{1}{2}\left(-\mathcal{F}(r)\dot{t}^2+\mathcal{F}^{-1}(r)\dot r^2+r^2\dot{\phi}^2\right).
\end{equation}
The equations of motion are obtained from
\begin{equation}\label{eq:lagrangian_equations}
    \frac{\md}{\md\lambda}\left(\frac{\partial\mathcal{L}}{\partial\dot{x}^\mu}\right)-\frac{\partial\mathcal{L}}{\partial x^\mu}=0.
\end{equation}
Along null geodesics, the following condition holds:
\begin{equation}\label{eq:null_condition}
    \mathcal{G}_{\mu\nu}\dot{x}^\mu\dot{x}^\nu=-\mathcal{F}(r)\dot{t}^2+\mathcal{F}^{-1}(r)\dot{r}^2+r^2\dot{\phi}^2\equiv 0.
\end{equation}
Notice that we consider null geodesic in the equatorial plane, i.e., $\theta=\frac{\pi}{2}$. 

\begin{figure}[ht]
    \centering
    \includegraphics[scale=0.4]{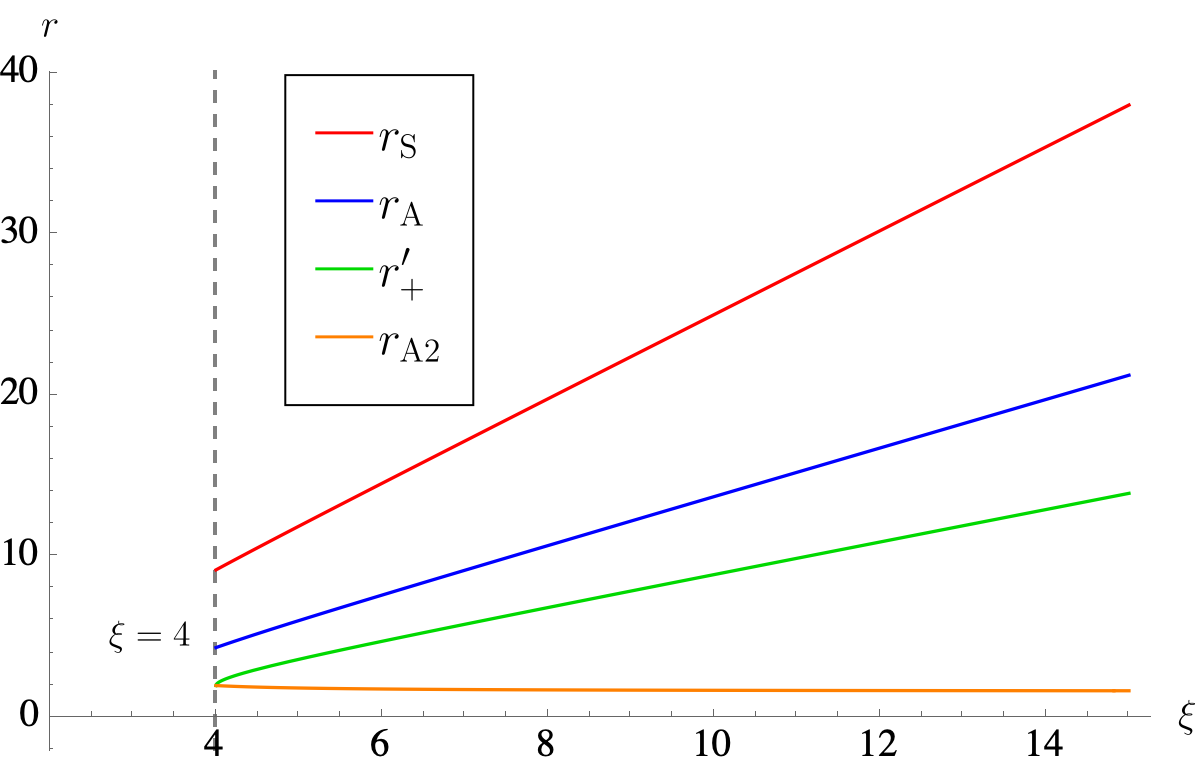}
    \caption{\raggedright Acoustic black hole shadow and associated radii as functions of $\xi$ at fixed $L=L_0$. Acoustic shadow radius $r_\text S$, acoustic sphere radius $r_\text A$, acoustic horizon $r_+'$, and the second largest acoustic sphere radius $r_{\text A2}$ are plotted starting from the critical value $\xi=4$.}
    \label{fig:rad_shad_sphere}
\end{figure}

The observer perceives themselves to be in Euclidean space \cite{Perlick:2021aok}, so the shadow radius corresponds to $r_{\text{S}}$ as determined by the tangent of the red geodesic, as shown in Fig.~\ref{fig:bh_shadow}. Moreover, the relation between $r_{\text{S}}$ and $r_{\text{O}}$ is
\begin{equation}\label{eq:tangent}
    \rsh=\tan\alpha\ r_{\text{O}}=\left.\frac{\sqrt{\mathcal G_{\phi\phi}}\md\phi}{\sqrt{\mathcal{G}_{rr}}\md r}\right|_{r_{\text O}}r_{\text O}=\sqrt{r_{\text{O}}^2\mathcal{F}(r_\text{O})}\left.\frac{\md\phi}{\md r}\right|_{r_\text O}r_{\text O}
\end{equation}
where $\frac{\md \phi}{\md r}$ can be obtained from Eq.~\eqref{eq:null_condition} as
\begin{equation}\label{eq:square_direv}
    \frac{\md \phi}{\md r}=\frac{\dot \phi}{\dot r}=\left(\mc{F}^2(r)\frac{\dot{t}^2}{\dot{\phi}^2}-r^2\mc F(r)\right)^{-\frac{1}{2}},
\end{equation}
where $\dot t$ and $\dot\phi$ are determined by the $t$ and $\phi$ components of the Lagrange equation \eqref{eq:lagrangian_equations}, respectively. These are related to two constants of motion along the red geodesic, expressed as
\begin{equation}\label{eq:constant}
    E=\mathcal{F}(r)\dot{t},\quad L=r^2\dot{\phi}.
\end{equation}
Hence $\frac{\md \phi}{\md r}$ can be expressed in terms of $E$ and $L$ as
\begin{equation}\label{eq:drdp}
    \frac{\md \phi}{\md r}=\left(\frac{E^2}{L^2}r^2-\mathcal{F}(r)\right)^{-\frac{1}{2}}r^{-1}.
\end{equation}
Substituting this result into Eq.~\eqref{eq:tangent} yields
\begin{equation}\label{eq:rS}
    r_{\text{S}}=\left(\frac{E^2}{L^2}\frac{r_\text{O}^2}{\mathcal F(r_\text O)}-1\right)^{-\frac{1}{2}}r_\text O,
\end{equation}
where $\frac{E}{L}$ is fixed by an additional property of the critical red geodesic: for such an orbit, the minimum distance to the acoustic black hole should be the radius of the acoustic sphere $r_\text{A}$, as shown in Fig.~\ref{fig:bh_shadow}, which describes the circular null geodesics around the acoustic black hole \cite{Perlick:2021aok}. This condition implies
\begin{equation}\label{eq:turning_point}
    \left.\frac{\md r}{\md \phi}\right|_{r_{\text{A}}}=\left(\frac{E^2}{L^2}r_{\text{A}}^2-\mathcal{F}(r_{\text{A}})\right)^{\frac{1}{2}}r_{\text{A}}=0,
\end{equation}
where $r_{\text{A}}$ denotes the radius of the acoustic sphere. Then substituting  
\begin{equation}\label{eq:relat_elr}
    \left(\frac{E}{L}\right)^2=\frac{\mathcal{F}(r_{\text{A}})}{r_{\text{A}}^2}
\end{equation}
into Eq.~\eqref{eq:rS} gives
\begin{equation}
    r_{\text S}=\left(\frac{\mathcal F(r_{\text A})}{\mathcal F(r_{\text O})}\frac{r^2_{\text O}}{r^2_{\text A}}-1\right)^{-\frac{1}{2}}r_{\text O}=\left(\frac{1}{\mathcal{F}(r_{\text{O}})}\frac{\mathcal{F}(r_{\text{A}})}{r_{\text{A}}^2}-\frac{1}{r_{\text{O}}^2}\right)^{-\frac{1}{2}}.
\end{equation}
For an observer at infinity, $r_{\text{O}}\rightarrow\infty$, $\mathcal{F}(r_{\text{O}})\rightarrow1$. As a result,
\begin{equation}\label{eq:shadow}
    r_{\text{S}}\rightarrow\left(\frac{\mathcal{F}(r_{\text{A}})}{r_{\text{A}}^2}\right)^{-\frac{1}{2}}=\frac{r_{\text{A}}}{\sqrt{\mathcal{F}(r_{\text{A}})}}.
\end{equation}
The final quantity to determine is the radius of the acoustic sphere $r_{\text{A}}$. This can be obtained by evaluating $\frac{\md^2 r}{\md\phi^2}$ of the acoustic sphere, which is done by differentiating $\frac{\md r}{\md\phi}$. On acoustic sphere, $r$ is constant, so this second derivative must vanish, i.e., 
\begin{equation}
   \left.\frac{\md^2 r}{\md\phi^2}\right|_{r_\text{A}}= \left[\left(\frac{(E)^2}{(L)^2}r^2-\mathcal{F}(r)\right)^{\frac{1}{2}}
    +\frac{1}{2}\left(\frac{(E)^2}{(L)^2}r^2-\mathcal{F}(r)\right)^{-\frac{1}{2}}
    \left(\frac{(E)^2}{(L)^2}2r-\mathcal{F}^\prime(r)\right)r\right]\frac{\md r}{\md\phi}=0,
\end{equation}
where $E$ and $L$ clearly satisfy Eq.~\eqref{eq:turning_point} as well, and $\mathcal{F}^\prime(r)=\frac{\md\mathcal{F}(r)}{\md r}$. Therefore, $r_{\text{A}}$ is the solution of 
\begin{equation}\label{eq:photon_sphere}
    2\mathcal{F}(r_{\text{A}})-r_{\text{A}}\mathcal{F}'(r_{\text{A}})=0,
\end{equation}
which can be solved numerically using Mathematica. In general black hole spacetimes, multiple photon (acoustic) spheres exist, complicating the analysis. For the acoustic Hayward black hole, however, only one acoustic sphere exists outside the event horizon. Substituting this solution into Eq.~\eqref{eq:shadow} yields the radius of black hole shadow $r_{\text{S}}$ as illustrated in Fig.~\ref{fig:rad_shad_sphere}, in which we plot the shadow radius $r_\text S$, the largest acoustic sphere radius $r_{\text A}$, the horizon radius $r_+'$ and the second largest acoustic sphere radius $r_{\text A2}$ as functions of the parameter $\xi$. It can be observed that the largest acoustic sphere lies outside the event horizon, while the second largest one lies inside, as earlier mentioned. Moreover, starting from $\xi=4$, as $\xi$ increases, $r_\text S$, $r_\text A$ and $r_+'$ all increase monotonically, with the order $r_\text S>r_\text A>r_+'$ maintained throughout.

\begin{figure}[ht]
    \centering
    \begin{subfigure}{0.47\textwidth}
        \centering
        \includegraphics[width=\textwidth]{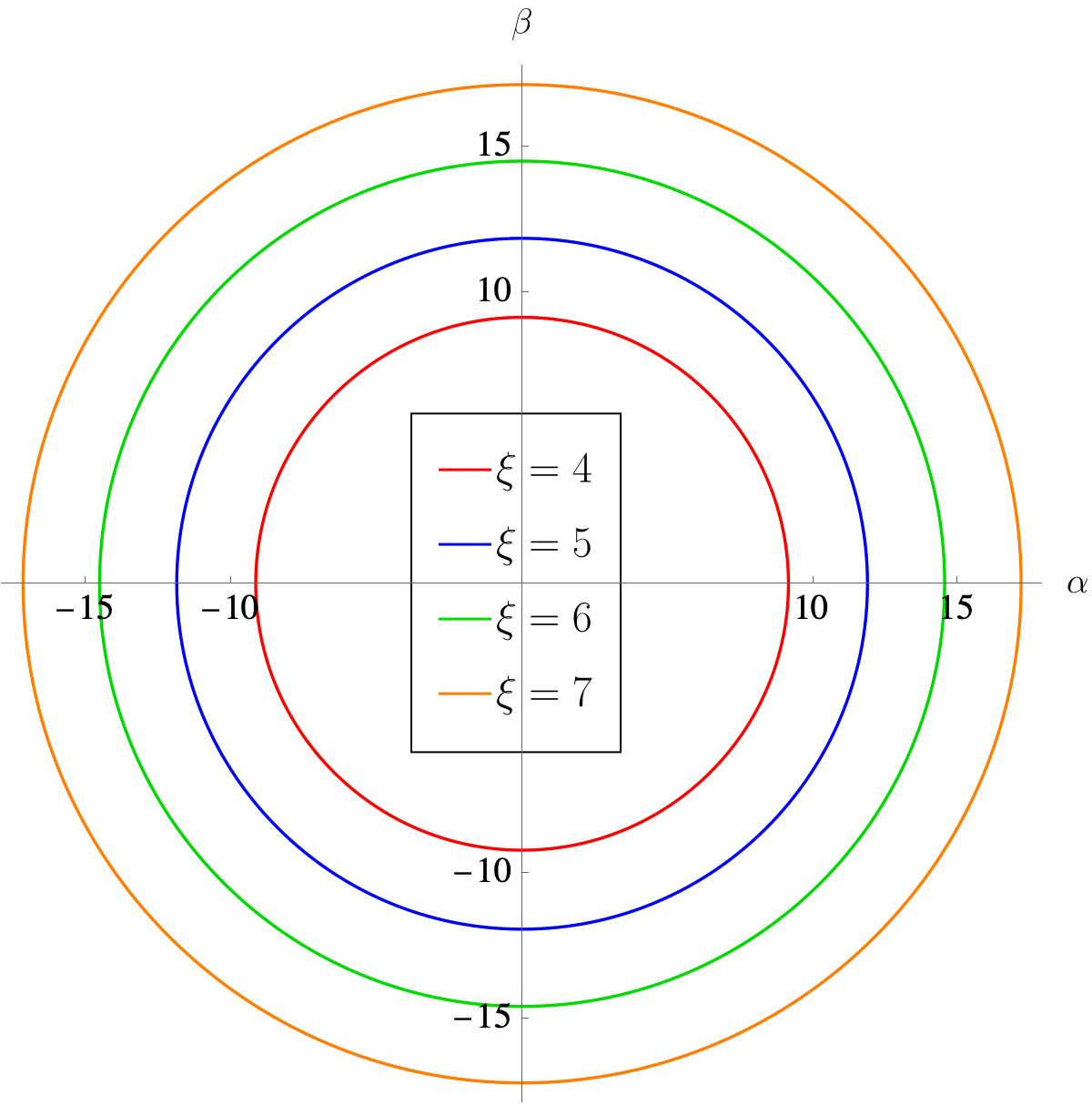}
    \end{subfigure}
    \hspace{0.5cm}
    \begin{subfigure}{0.47\textwidth}
    \centering
    \includegraphics[width=\textwidth]{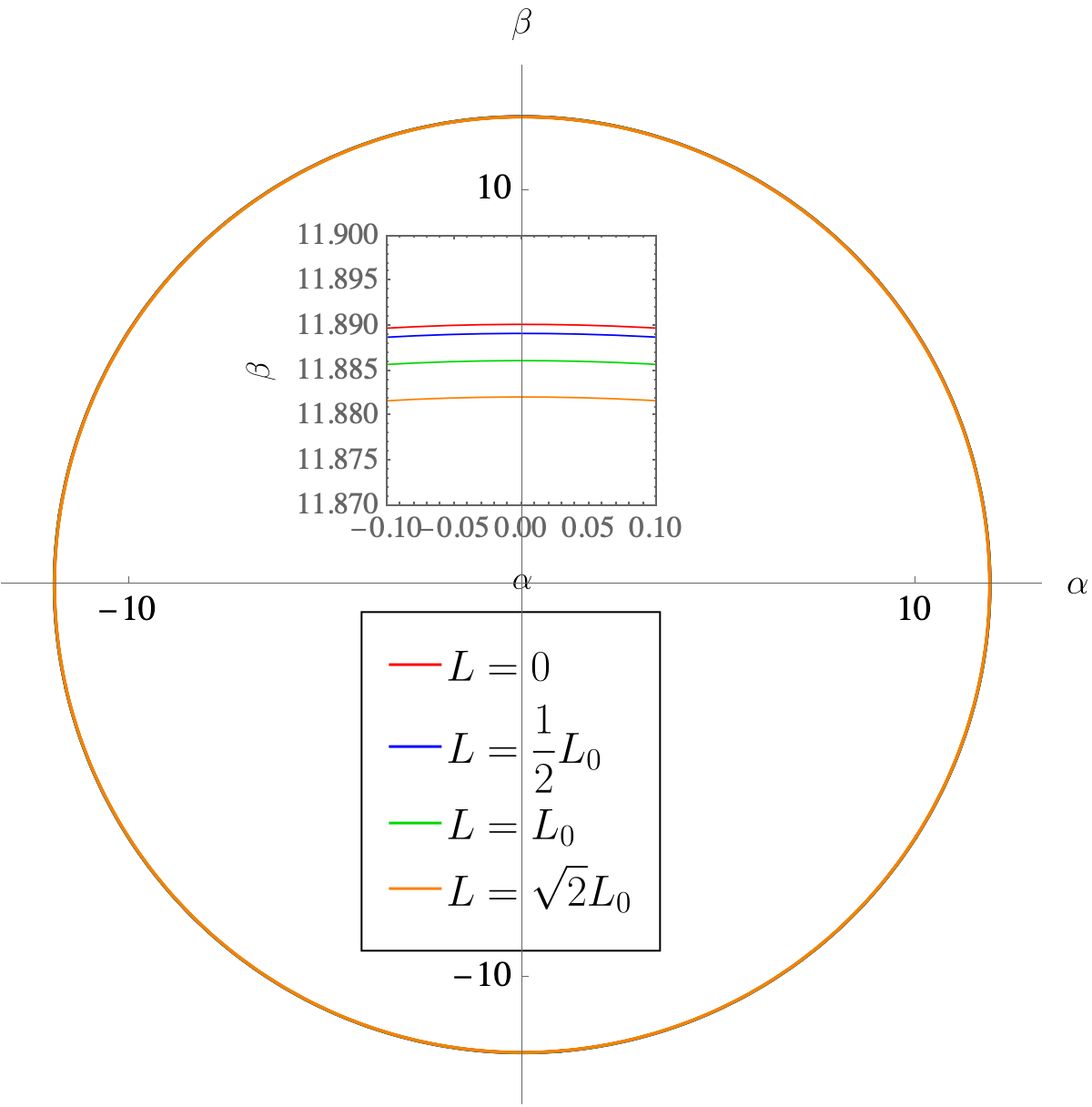}
    \end{subfigure}
    \caption{\raggedright Acoustic shadow patterns for different values of $\xi$ and $L$. Left panel: $L=L_0$ is fixed. Right panel: $\xi =5 $ is fixed, with an inserted magnified view of a local region.}
    \label{fig:sha_pic}
\end{figure}

Given that the acoustic Hayward black hole is spherically symmetric, once the shadow radius $r_\text S$ is obtained, the approximate image of the black hole shadow can be constructed. By defining $r_\text S=\sqrt{\alpha^2+\beta^2}$, the shadows are plotted in the $(\alpha,\beta)$ plane for different parameter choices, as shown in Fig.~\ref{fig:sha_pic}. In the left panel, with $L=L_0$ fixed, we plot the acoustic shadow for different values of tuning parameter $\xi=4,5,6,7$. As $\xi$ increases, the black hole shadow also expands. In the right panel, with $\xi=5$ fixed, we plot the acoustic shadow for different values of the Hayward parameter $L=0,\frac{L_0}{2}, L_0,\sqrt{2}L_0$. The variation in the acoustic shadow with respect to $L$ is relatively small. To highlight the subtle trend, an inset subfigure is included, which reveals a slight decrease in the acoustic shadow as $L$ increases. The variation of the acoustic shadow with respect to $\xi$ follows the same pattern as in the acoustic Schwarzschild and charged black hole cases \cite{Guo:2020blq,Ling:2021vgk}, while its dependence on $L$ is consistent with that observed for the Hayward black hole \cite{Vagnozzi:2022moj}.

\section{Covariant scalar field equation and the effective potential}\label{sec:Cov_eq}
Another important property of black holes is their influence on the fields in their vicinity. For instance, black holes can affect the propagation and decay of the fields, as manifested in the behaviors of QNMs, and can also influence the grey-body factor and energy emission rate in Hawking radiation. All these phenomena can be studied by solving the test fields near the black hole. Therefore, it is first necessary to derive the covariant field equations in the black hole background. Then, for different physical processes, appropriate boundary conditions are chosen, and the resulting equations are solved accordingly. For acoustic black holes, the fluctuation of phase $\theta_1$ in Eq.~\eqref{eq:covariant_eq_theta} is a massless scalar field $\psi$, i.e., the phonon, satisfying
\begin{equation}\label{eq:covariant_eq_psi}
    \frac{1}{\sqrt{-\mc G}}\partial_\mu\left({\sqrt{-\mc{G}}\mc G^{\mu\nu}\partial_\nu\psi}\right)=0.
\end{equation}
Following the standard separation of variables, we write
\begin{equation}
    \psi(t,r,\theta)=\sum_{l,m}e^{-i\omega t}\frac{\Psi(r)}{r}Y_{l,m}(\theta,\phi),
\end{equation}
which forms a complete basis. Substituting this ansatz into Eq.~\eqref{eq:covariant_eq_psi} and summing over all components yields
\begin{equation}
    \mathcal{F}^2\frac{\md^2\Psi}{\md r^2}+\mathcal{F}\mathcal{F}'\frac{\md\Psi}{\md r}+\omega^2\Psi-\mc F\left(\frac{l(l+1)}{r^2}+\frac{\mc F'}{r}\right)\Psi=0,
\end{equation}
where a prime denotes a derivative with respect to $r$. Introducing the tortoise coordinate defined by $\md r_*=\mc F^{-1}\md r$, the first two terms combine into $\frac{\md^2\Psi}{\md r_*^2}$. The equation then simplifies to 
\begin{equation}
    \frac{\md^2\Psi}{\md r_*^2}+\omega^2\Psi-\mc F\left(\frac{l(l+1)}{r^2}+\frac{\mc F'}{r}\right)\Psi=0.
\end{equation}
Defining the effective potential as
\begin{equation}\label{eq:eff_pont}
    V(r)=\mc F\left(\frac{l(l+1)}{r^2}+\frac{\mc F'}{r}\right),
\end{equation}
the covariant equation reduces to a Schr\"{o}dinger-like equation
\begin{equation}\label{eq:schro}
    \frac{\md^2\Psi}{\md r_*^2}+\left(\omega^2-V(r)\right)\Psi=0.
\end{equation}

\begin{figure}[ht]
    \centering
    \begin{subfigure}{0.47\textwidth}
        \centering
        \includegraphics[width=\textwidth]{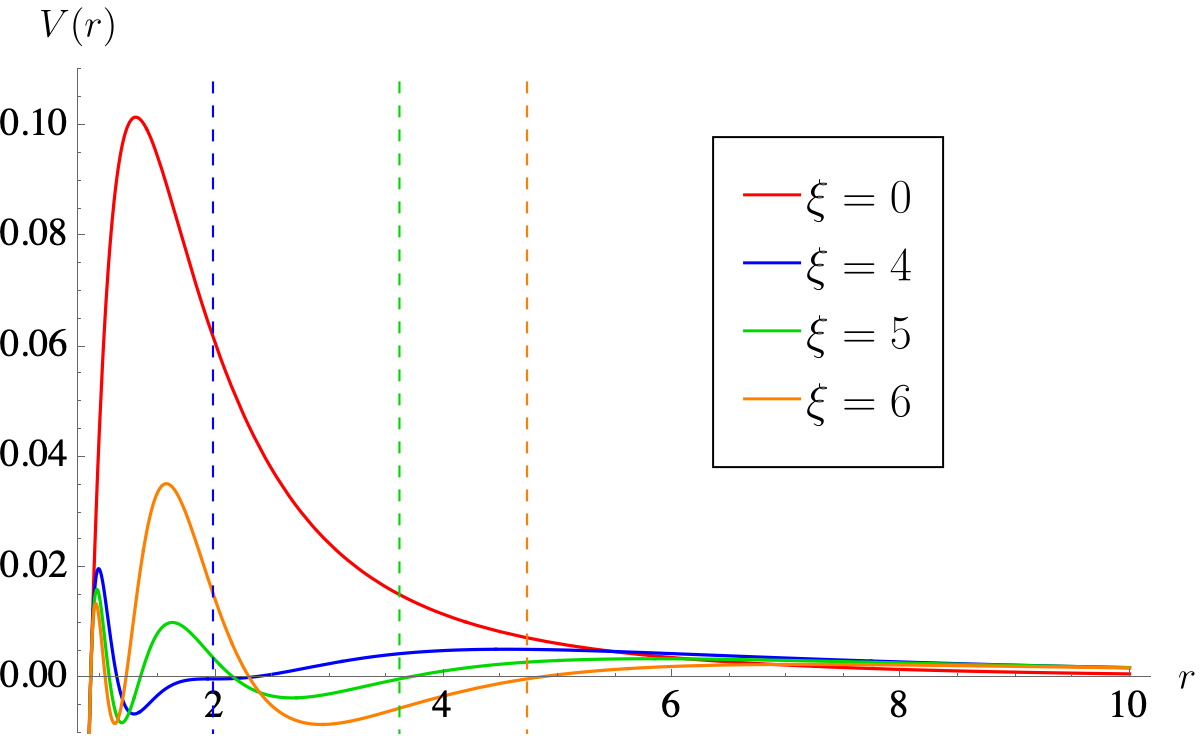}
    \end{subfigure}
    \hspace{0.5cm}
    \begin{subfigure}{0.47\textwidth}
    \centering
    \includegraphics[width=\textwidth]{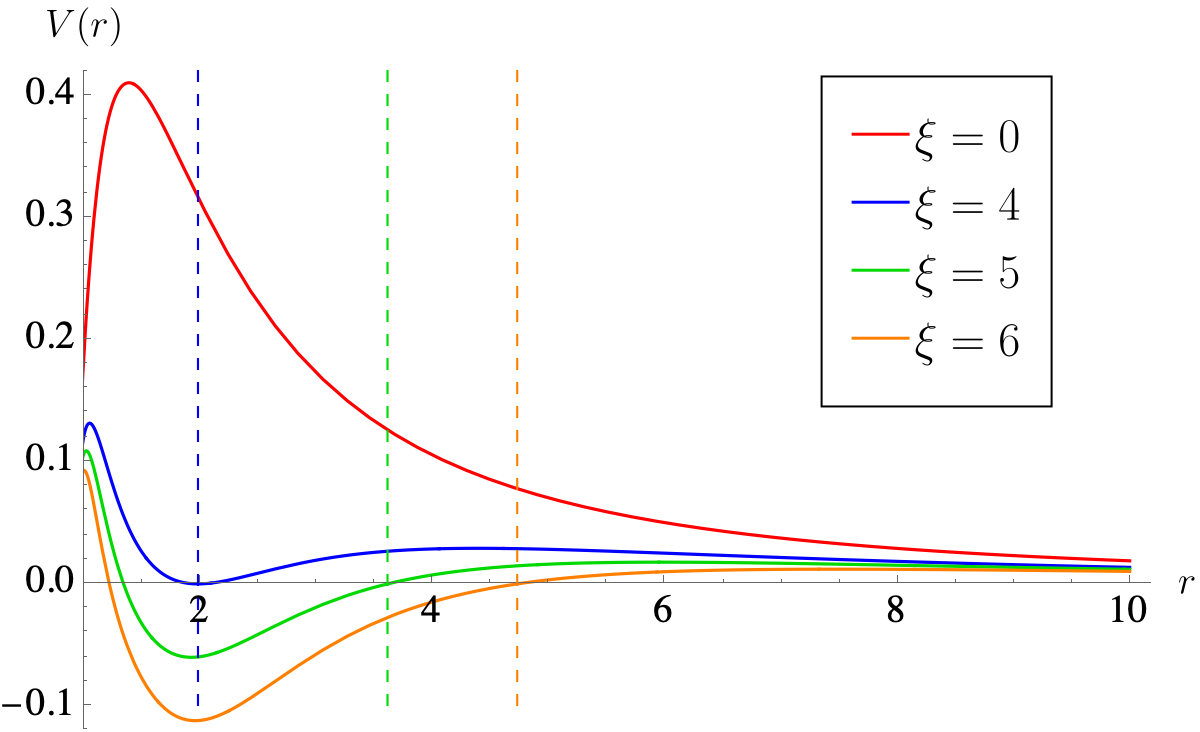}
    \end{subfigure}
    \caption{\raggedright Effective potential $V(r)$ as a function of the radial coordinate $r$ for several values of the tuning parameter $\xi$, with $L=L_0$ fixed. Left panel: $l=0$ is fixed. Right panel: $l=1$ is fixed. }
    \label{fig:poten_xi}
\end{figure}

\begin{figure}[ht]
    \centering
    \begin{subfigure}{0.47\textwidth}
        \centering
        \includegraphics[width=\textwidth]{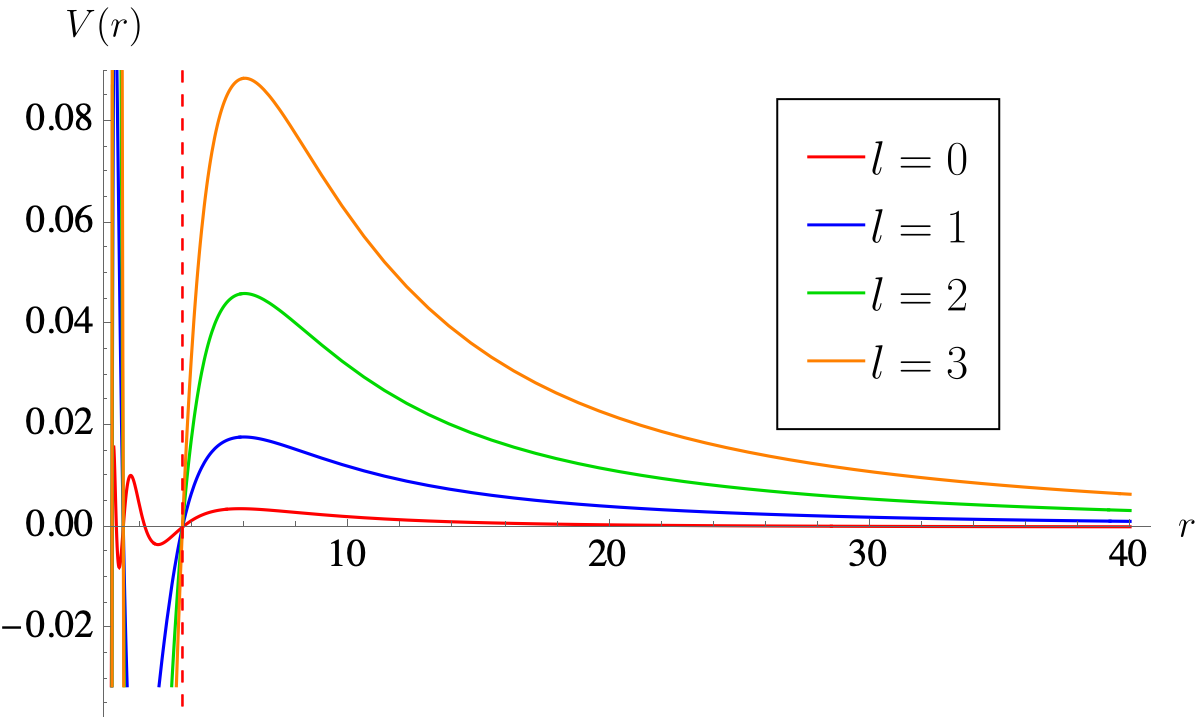}
    \end{subfigure}
    \hspace{0.5cm}
    \begin{subfigure}{0.47\textwidth}
    \centering
    \includegraphics[width=\textwidth]{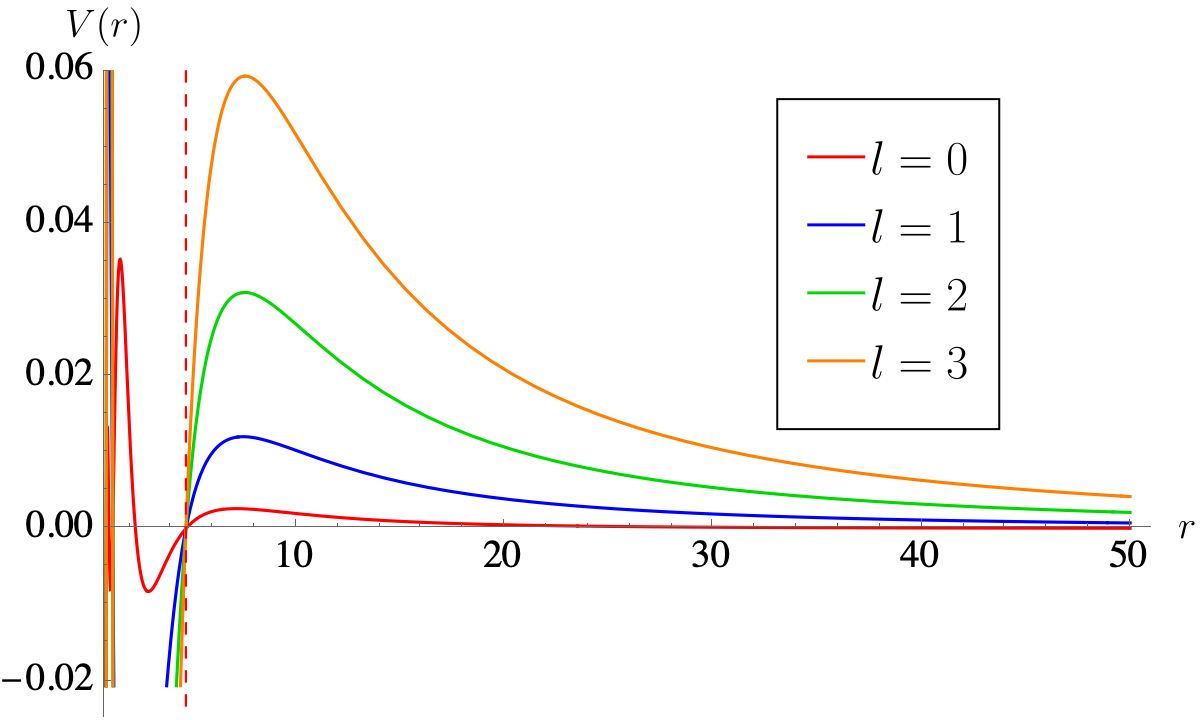}
    \end{subfigure}
    \caption{\raggedright Effective potential $V(r)$ as a function of the radial coordinate $r$ for several values of angular momentum number $l$, with $L=L_0$ fixed. Left panel: $\xi=5$ is fixed. Right panel: $\xi=6$ is fixed. }
    \label{fig:poten_l}
\end{figure}

The influence of a black hole on a scalar field is largely determined by the form of the effective potential $V(r)$, which is shown as a function of $r$ in Fig.~\ref{fig:poten_xi} and Fig.~\ref{fig:poten_l}, with the Hayward parameter fixed as $L=L_0$ in all cases. In Fig.~\ref{fig:poten_xi}, the angular quantum number is set to $l=0$ in the left panel and $l=1$ in the right panel, allowing a comparative study of the effective potential $V(r)$ for different values of the tuning parameter $\xi=0,4,5,6$, in which $\xi=0$ represents Hayward black hole. Meanwhile, in Fig.~\ref{fig:poten_l}, the tuning parameter is fixed as $\xi=5$ in the left panel and $\xi=6$ in the right panel, illustrating the dependence of the effective potential on the angular quantum number $l$.

In both figures, the effective potential vanishes at the acoustic horizon, marked by the vertical dashed line, exhibits a barrier-like profile outside the horizon, and asymptotically tends to zero at large $r$. In Fig.~\ref{fig:poten_xi}, as $\xi$ increases, the peak height of the potential barrier decreases. Moreover, the location of the zero potential, which corresponds to the acoustic horizon, shifts outward to larger radii, in agreement with earlier analysis. In Fig.~\ref{fig:poten_l}, the peak of the effective potential increases with $l$, while the position of zero potential remains unchanged, as the radius of the acoustic horizon is independent of the angular quantum number $l$. The influence of the parameter $\xi$ and angular momentum number $l$ on the effective potential discussed above will manifest in the subsequent analyses of QNMs and Hawking radiation.

\section{Quasinormal mode frequencies}\label{sec:QNMs}
Once the Schr\"{o}dinger-like equation \eqref{eq:schro} and effective potential Eq.~\eqref{eq:eff_pont} are established, both the QNMs and Hawking radiation can be investigated. We first consider the computation of QNMs. The physical process associated with QNMs can be understood as a ``knock'' or a perturbation outside the black hole, and the QNM, which is characterized by the complex frequency $\omega$, governs the propagation and decay of this perturbation. In the wave function $\psi$ \eqref{eq:covariant_eq_psi}, the temporal dependence takes the form $e^{-i\omega t}=e^{\text{Im}(\omega )t}e^{-i\text{Re}(\omega)t}$, which implies that:
\begin{itemize}
    \item The imaginary part $\text{Im}(\omega)$ determines the decay (or growth) rate, with $\text{Im}(\omega)<0$ corresponding to a decaying mode;
    \item The real part $\text{Re}(\omega)$ gives the oscillation frequency, i.e. the propagating mode, and is typically taken as positive, $\text{Re}(\omega)>0$. 
\end{itemize}
Correspondingly, the appropriate boundary conditions for the wave equation are:
\begin{itemize}
    \item At the event horizon ($r_*\rightarrow -\infty$), purely ingoing waves, $\Psi\sim e^{-i\omega r_*}$;
    \item At spatial infinity ($r_*\rightarrow +\infty$), purely outgoing waves, $\Psi\sim e^{+i\omega r_*}$.
\end{itemize}
Due to the complexity of the effective potential \eqref{eq:eff_pont}, we employ the WKB method  \cite{konoplya2019higher} to compute the QNMs numerically, and validate the results using the asymptotic iteration method (AIM) \cite{Cho:2009cj,Cho:2011sf}. The idea of WKB method is to perform a WKB expansion of $\Psi$ near the event horizon and at spatial infinity, i.e., for $r_*\rightarrow\pm\infty$ \cite{Konoplya:2011qq,Berti:2009kk}. A Taylor expansion for $Q(r_*)=\omega^2-V(r)$ is carried out around the peak of the effective potential, which allows $\Psi$ to be expressed in terms of parabolic cylinder functions $D_\nu$, where
\begin{equation}\label{eq:nuWKB}
    \nu+\frac{1}{2}=\frac{-iQ_0}{\sqrt{2Q_0''}}-\sum_{i=2}^N\Lambda_i,
\end{equation}
where $Q_0=Q(r_{*0})$ is the maximum of $Q(r_*)$, and $\Lambda_i$ are higher-order correction terms. Explicit expressions for $\Lambda_2,\Lambda_3$ can be found in \cite{Iyer:1986np}, for $\Lambda_4-\Lambda_6$ in \cite{Konoplya:2003ii}, and for $\Lambda_7-\Lambda_{13}$ in \cite{Matyjasek:2017psv}. The integer $N$ is referred to as the  WKB order. By asymptotically analyzing this solution and matching it to the WKB expansions, the scattering matrix ($S$ matrix) can be derived \cite{Konoplya:2011qq}. The boundary conditions of QNMs require that $\Gamma(-\nu)$ in the $S$ matrix be singular. As a result, $\nu$ must be an integer. Eq.~\eqref{eq:nuWKB} therefore gives the WKB formula for $\omega$: 
\begin{equation}\label{eq:wkbw}
    \frac{i(\omega^2-V_0)}{\sqrt{-2V_0''}}-\sum_{i=2}^N\Lambda_i=n+\frac{1}{2},\quad n=0,1,2,\cdots
\end{equation}
where $n$ is called the overtone number. We use the Mathematica code provided in \cite{konoplya2019higher} to compute the QNMs. The results converge well in the 6th-order WKB approximation, nevertheless, we carry out the calculation up to the 9th-order for improved accuracy.

The results for tuning parameter $\xi\leq 10$ are shown in Tabs.~\ref{tab:n0l0},\ref{tab:n0l1},\ref{tab:n1l1}, where Tab.~\ref{tab:n0l0} corresponds to the mode $n=0, l=0$, Tab.~\ref{tab:n0l1} to $n=0, l=1$, and Tab.~\ref{tab:n1l1} to $n=1, l=1$. In each table, we present the QNMs computed using the 9th-order WKB method and AIM.

\begin{table}[h]
    \centering
     \renewcommand{\arraystretch}{1.1} 
    \setlength{\tabcolsep}{18pt}
    \begin{tabular}{c |c |c}
    \hline
    $\xi$ & 9th-order WKB & AIM \\
    \hline
    0  & $0.229004-0.194160i$ &  -  \\
    4  & $0.056704-0.038036i$ &  -  \\
    5  & $0.046820-0.033090i$ &  $0.046997-0.033164i$\\
    6  & $0.039084-0.029403i$ &  $0.039107-0.029420i$\\
    7  & $0.033356-0.026193i$ &  $0.033321-0.026143i$\\
    8  & $0.029382-0.023358i$ &  $0.028989-0.023417i$\\
    9  & $0.025671-0.021313i$ &  $0.025642-0.021161i$\\
    10 & $0.022958-0.019425i$ &  $0.022983-0.019279i$\\
    \hline
    \end{tabular}
    \caption{QNM frequencies for the acoustic Hayward black hole, computed for $0 \leq \xi \leq 10$ in the mode $n=0, l=0$. Left column: 9th-order WKB results. Right column: AIM validation.}
    \label{tab:n0l0}
\end{table}

\begin{table}[htbp]
    \centering
     \renewcommand{\arraystretch}{1.1} 
    \setlength{\tabcolsep}{18pt}
    \begin{tabular}{c |c |c}
    \hline
    $\xi$ & 9th-order WKB & AIM \\
    \hline
    0  & $0.600961-0.183503i$ &  -  \\
    4  & $0.164347-0.034759i$ &  -  \\
    5  & $0.127850-0.031791i$ &  $0.127850-0.031791i$\\
    6  & $0.104689-0.028027i$ &  $0.104689-0.028027i$\\
    7  & $0.088693-0.024794i$ &  $0.088693-0.024794i$\\
    8  & $0.076965-0.022141i$ &  $0.076965-0.022141i$\\
    9  & $0.067990-0.019965i$ &  $0.067990-0.019964i$\\
    10 & $0.060896-0.018160i$ &  $0.060896-0.018160i$\\
    \hline
    \end{tabular}
    \caption{\raggedright QNM frequencies for the acoustic Hayward black hole, computed for $0 \leq \xi \leq 10$ in the mode $n=0, l=1$. Left column: 9th-order WKB results. Right column: AIM validation.}
    \label{tab:n0l1}
\end{table}

\begin{table}[htbp]
    \centering
    \renewcommand{\arraystretch}{1.1} 
    \setlength{\tabcolsep}{18pt}
    \begin{tabular}{c |c |c}
    \hline
    $\xi$ & 9th-order WKB & AIM \\
    \hline
    0  & $0.549237-0.569090i$ &  -  \\
    4  & $0.153030-0.106792i$ &  - \\
    5  & $0.121258-0.097278i$ &  $0.121258-0.097278i$\\
    6  & $0.098970-0.086232i$ &  $0.098966-0.086231i$\\
    7  & $0.083434-0.076551i$ &  $0.083427-0.076548i$\\
    8  & $0.072068-0.068514i$ &  $0.072074-0.068519i$\\
    9  & $0.063427-0.061885i$ &  $0.063432-0.061889i$\\
    10 & $0.056634-0.056364i$ &  $0.566388-0.056368i$\\
    \hline
    \end{tabular}
    \caption{\raggedright QNM frequencies for the acoustic Hayward black hole, computed for $0 \leq \xi \leq 10$ in the mode $n=1, l=1$. Left column: 9th-order WKB results. Right column: AIM validation.}
    \label{tab:n1l1}
\end{table}

\begin{figure}[ht]
    \centering
    \begin{subfigure}{0.47\textwidth}
        \centering
        \includegraphics[width=\textwidth]{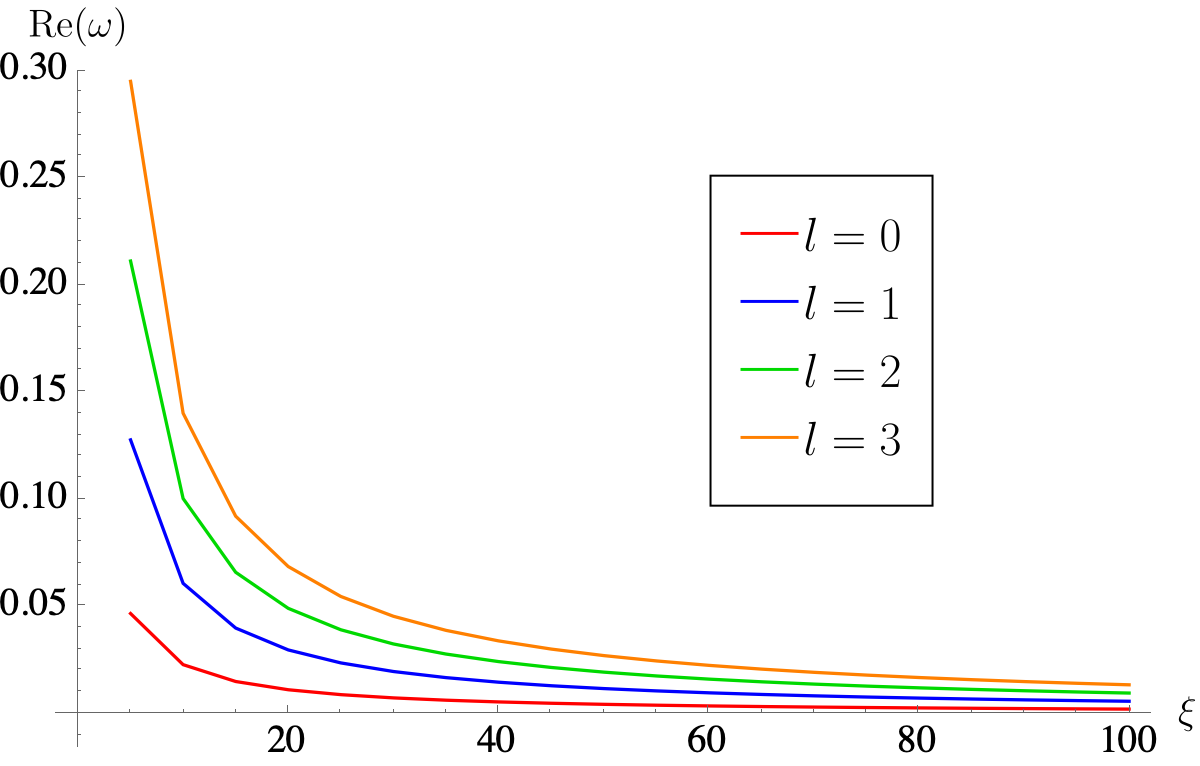}
    \end{subfigure}
    \hspace{0.5cm}
    \begin{subfigure}{0.47\textwidth}
    \centering
    \includegraphics[width=\textwidth]{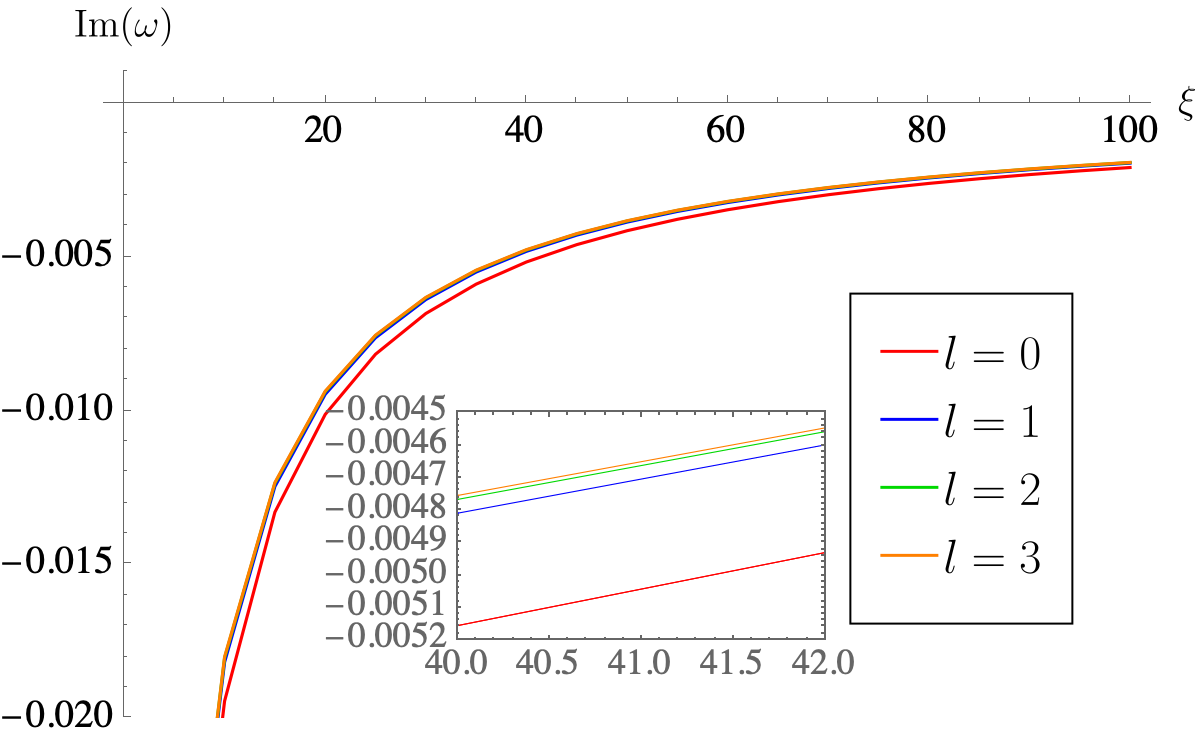}
    \end{subfigure}
    \caption{\raggedright QNM frequencies as functions of $\xi$ for different angular numbers $l$ with fixed $n=0$: real part (left panel) and imaginary part (right panel). A magnified view is shown in the inset of the right panel.}
    \label{fig:qnm_l}
\end{figure}

\begin{figure}[ht]
    \centering
    \begin{subfigure}{0.47\textwidth}
        \centering
        \includegraphics[width=\textwidth]{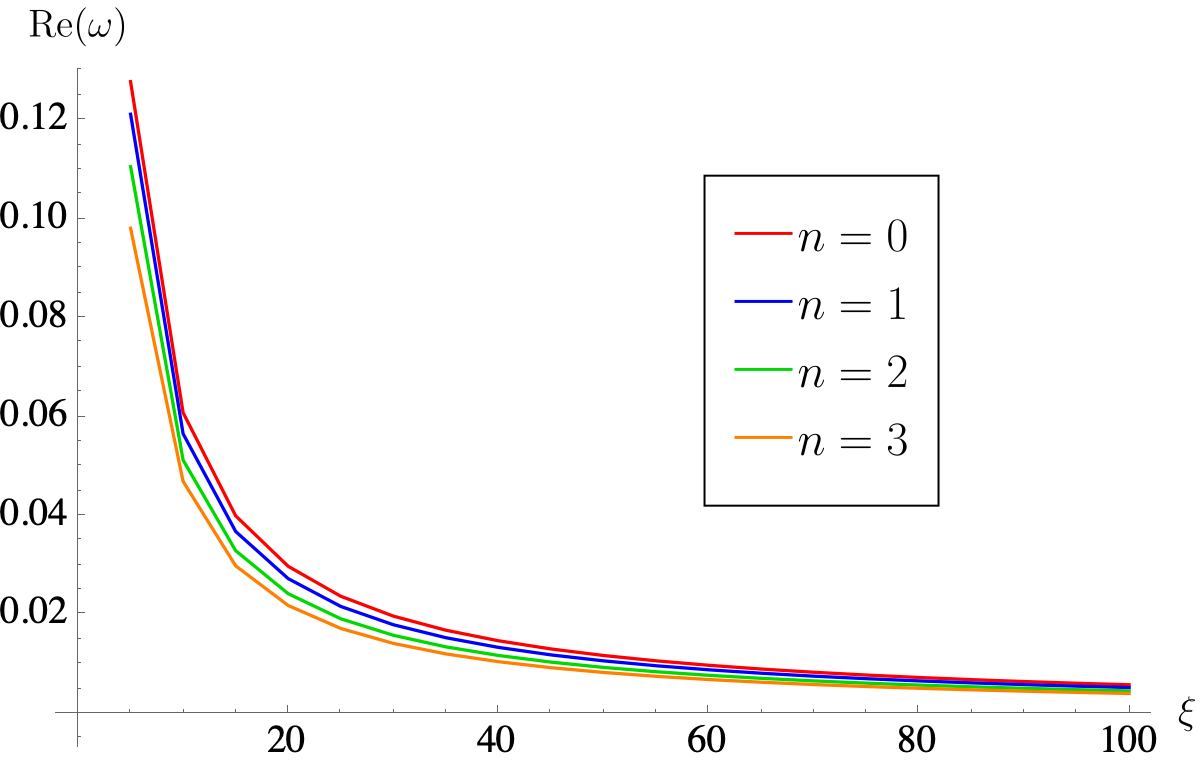}
    \end{subfigure}
    \hspace{0.5cm}
    \begin{subfigure}{0.47\textwidth}
    \centering
    \includegraphics[width=\textwidth]{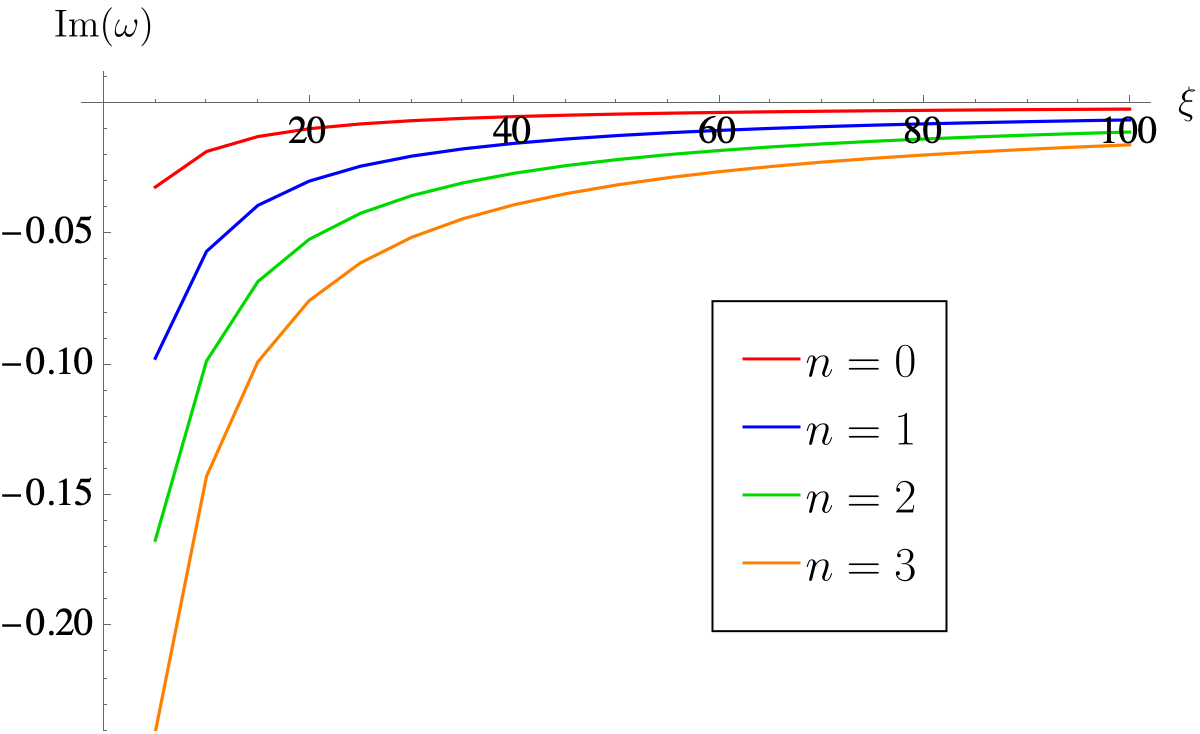}
    \end{subfigure}
    \caption{\raggedright QNM frequencies as functions of $\xi$ for different overtone numbers $n$ with fixed $l=1$: real part (left panel) and imaginary part (right panel).}
    \label{fig:qnm_n}
\end{figure}

The universal properties of the QNM frequencies can be summarized as follows:
\begin{itemize}
    \item In all modes, the real part satisfies $\text{Re}(\omega)>0$, and the imaginary part satisfies $\text{Im}(\omega)<0$, indicating that the acoustic Hayward black hole remains stable under the perturbation for small $\xi$, as we established earlier. This is consistent with the behavior in Hayward black hole. \cite{Flachi:2012nv,lopez2021quasinormal}
    \item The QNM amplitudes of acoustic Hayward black hole with $\xi\geq 4$ are significantly smaller than those of the Hayward black hole with $\xi=0$. This indicates that the QNMs signals are weaker in acoustic black hole compared to astrophysical black hole, which is consistent with previous findings \cite{Guo:2020blq,Ling:2021vgk}.
    \item As $\xi$ increases, the real part and the magnitude of the imaginary part of QNMs decrease. It means that both the oscillation frequency and the decay rate decrease, which can be attributed to the suppression of the effective potential and the alteration of the spacetime geometry as $\xi$ rises, as shown in Fig.~\ref{fig:poten_xi} and Eq.~\eqref{eq:Fr}. The decrease persists for $\xi>10$, as will be shown in Figs.~\ref{fig:qnm_n} and \ref{fig:qnm_l}. 
    \item In each table, the results obtained by the AIM and the WKB method show only minor discrepancies, confirming the accuracy of the WKB calculations.
\end{itemize}

In addition to presenting the values of QNMs for $\xi\geq 10$, we further investigate their behavior for larger $\xi$ at different overtone numbers $n$ and angular momentum numbers $l$, as illustrated in Figs.~\ref{fig:qnm_l} and \ref{fig:qnm_n}. In Fig.~\ref{fig:qnm_l}, with $n$ fixed at $0$, the dependence of the QNMs on $\xi$ is plotted for $l=0,1,2,3$. Conversely, Fig.~\ref{fig:qnm_n} presents the evolution of the QNMs with $\xi$ for $n=0,1,2,3$, while keeping $l=1$ fixed. In both figures, the left and right panels display the real and imaginary parts of the QNMs, respectively.

The following conclusions can be drawn from these figures:
\begin{itemize}
    \item In Figs.~\ref{fig:qnm_l} and \ref{fig:qnm_n}, it is evident that for all considered combinations of $n$ and $l$, the real part of the QNMs is consistently positive, while the imaginary part is negative. This indicates the stability of the acoustic Hayward black hole. Moreover, both the real and imaginary parts approach zero as $\xi$ increases, which is consistent with earlier discussions.
    \item In Fig.~\ref{fig:qnm_l}, as the angular momentum number $l$ increases, the real part frequency increases and the magnitude of imaginary part of the frequency decreases. This trend is attributed to the corresponding reduction in the effective potential with higher $l$, as shown in Fig.~\ref{fig:poten_l}.
    \item In Fig.~\ref{fig:qnm_n}, as the overtone number $n$ increases, the real part $\text{Re}(\omega)$ decreases while the magnitude of $\text{Im}(\omega)$ grows. This indicates a reduction in oscillation frequency together with an enhanced damping rate, which is consistent with the behaviors of other acoustic black holes \cite{Guo:2020blq,Ling:2021vgk}.
\end{itemize}

\section{Analogue Hawking radiation}\label{sec:aHR}
\begin{figure}[ht]
    \centering
    \begin{subfigure}{0.47\textwidth}
        \centering
        \includegraphics[width=\textwidth]{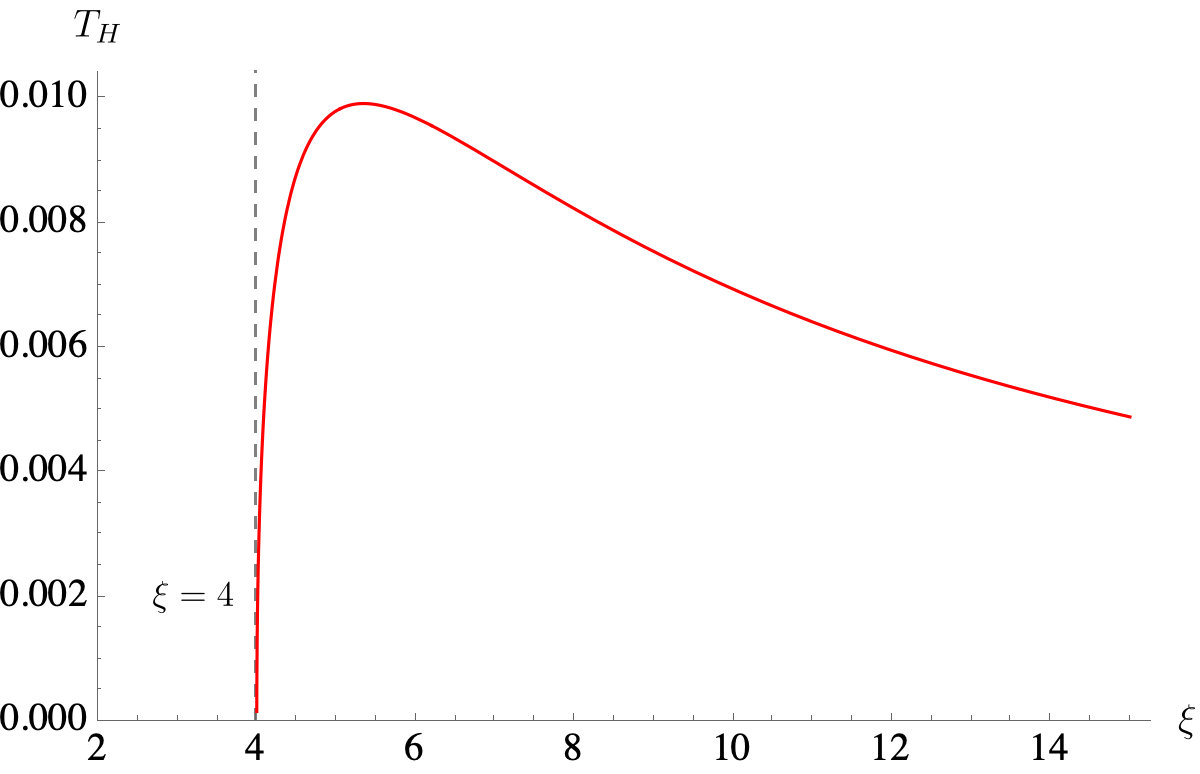}
    \end{subfigure}
    \hspace{0.5cm}
    \begin{subfigure}{0.47\textwidth}
    \centering
    \includegraphics[width=\textwidth]{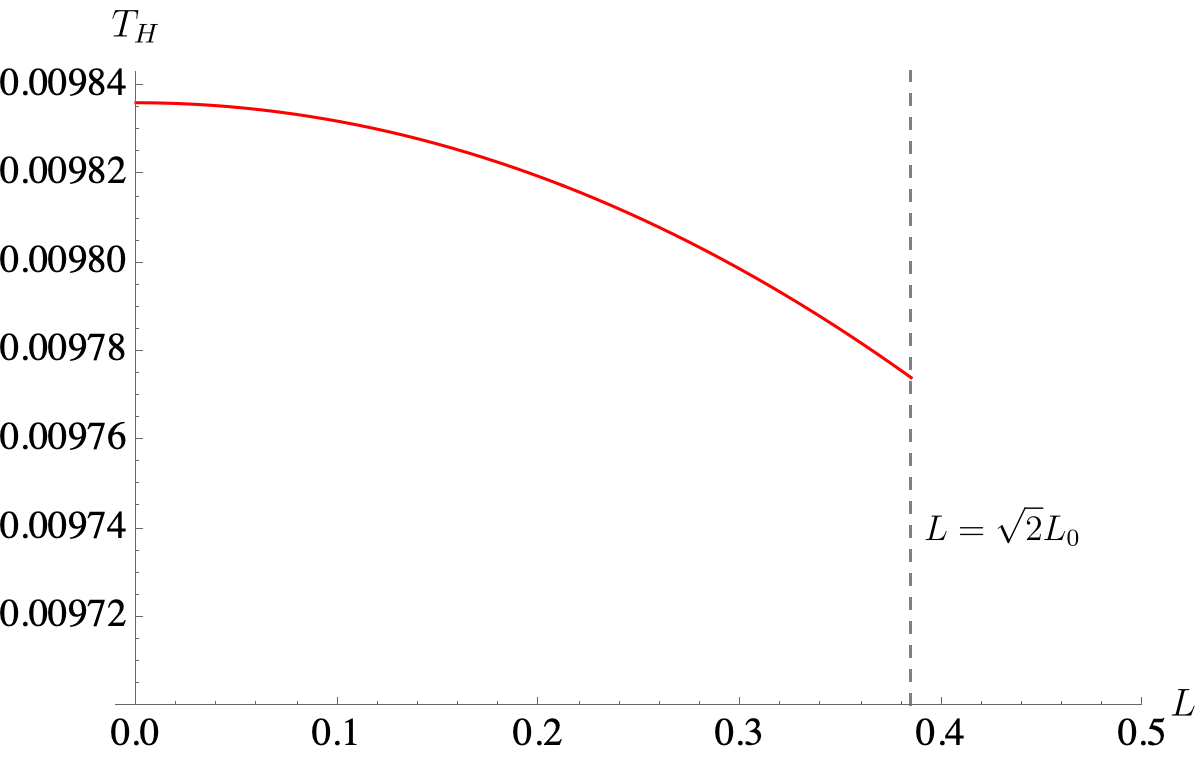}
    \end{subfigure}
    \caption{\raggedright Analogue Hawking temperature as a function of $\xi$ for $\xi \geq 4$ with $L = L_0$ (left panel), and as a function of $L$ for $0 \leq L \leq \sqrt{2}L_0$ with $\xi = 5$ (right panel).}
    \label{fig:haw_temp}
\end{figure}

In addition to QNMs, Hawking radiation can also be studied by solving the covariant field equations \eqref{eq:schro}. Similar to conventional black holes, acoustic black holes also emit analogue Hawking radiation, and correspondingly possess an analogue Hawking temperature \cite{Vieira:2014rva}. The underlying physical process of Hawking radiation can be understood as a wave originating near the horizon, which is then scattered by the effective potential. Part of the wave is reflected back, while another part tunnels through the potential barrier and reaches spatial infinity. Hence, Hawking radiation corresponds to a scattering problem with the following boundary conditions:
\begin{itemize}
    \item At event horizon ($r_*\rightarrow-\infty$), $\Psi=e^{-i\omega r_*}+R_l(\omega)e^{+i\omega r_*}$;
    \item At spatial infinity ($r_*\rightarrow+\infty$), $\Psi=T_l(\omega)e^{-i\omega r_*}$.
\end{itemize}
Here $R_l(\omega)$ and $T_l(\omega)$ denote the reflection and transmission coefficients, respectively, which depend on both $\omega$ and $l$ and satisfy the relation $|R_l(\omega)|^2+|T_l(\omega)|^2=1$. The grey-body factor $\mathcal{A}_l(\omega)$ for analogue Hawking radiation is exactly the transmission coefficient, giving $|\mathcal A_l(\omega)|^2=1-|R_l(\omega)|^2=|T_l(\omega)|^2$. Therefore, the grey-body factors can be obtained by imposing the scattering boundary conditions on the previously mentioned $S$ matrix  \cite{Iyer:1986np,Konoplya:2011qq}, namely
\begin{equation}
    |\mathcal A_l(\omega)|^2=|T_l(\omega)|^2=S_{21}^{-1}=\left(1+e^{2\pi i(n_l(\omega)+\frac{1}{2})}\right),
\end{equation}
where the relation between the overtone number $n$ and $\omega$ can be obtained from Eq.~\eqref{eq:wkbw}. Consequently, the grey-body factors can also be computed numerically using the WKB method \cite{konoplya2019higher,Konoplya:2024lir}, which modifies the black-body radiation spectrum according to $\frac{\md E_{gb}}{\md t}=|\mathcal A_l|^2\frac{\md E_{bd}}{\md t}$. Consequently, the energy emission rate for analogue Hawking radiation reads \cite{Hawking:1975vcx,Page:1976ki}
\begin{equation}
    \frac{\md E}{\md t}=\sum_l (2l+1)|\mathcal A_l(\omega)|^2\frac{\omega}{e^{\frac{\omega}{T_H}}-1}\frac{\md \omega}{2\pi},
\end{equation}
where the analogue Hawking temperature is $T_H=-\frac{\mc F'(r_{+}')}{4\pi}$. Fig.~\ref{fig:haw_temp} presents the dependence of the analogue Hawking temperature on $\xi$ and $L$. In the left panel, with $L$ fixed to $L_0$, the temperature initially increases with $\xi$ for $\xi\geq4$, followed by a subsequent decrease. The initial rise is attributed to changes in the effective potential peak and the energy emission rate, while the subsequent decline results from the alteration of spacetime geometry. In the right panel, with $\xi$ fixed at $\xi=5$, the analogue Hawking temperature decreases as $L$ increases over the range $0\leq L\leq \sqrt{2}L_0$. This reduction likewise originates from the suppression of the energy emission rate, which will be discussed in detail later.

\begin{figure}[ht]
    \centering
    \begin{subfigure}{0.47\textwidth}
        \centering
        \includegraphics[width=\textwidth]{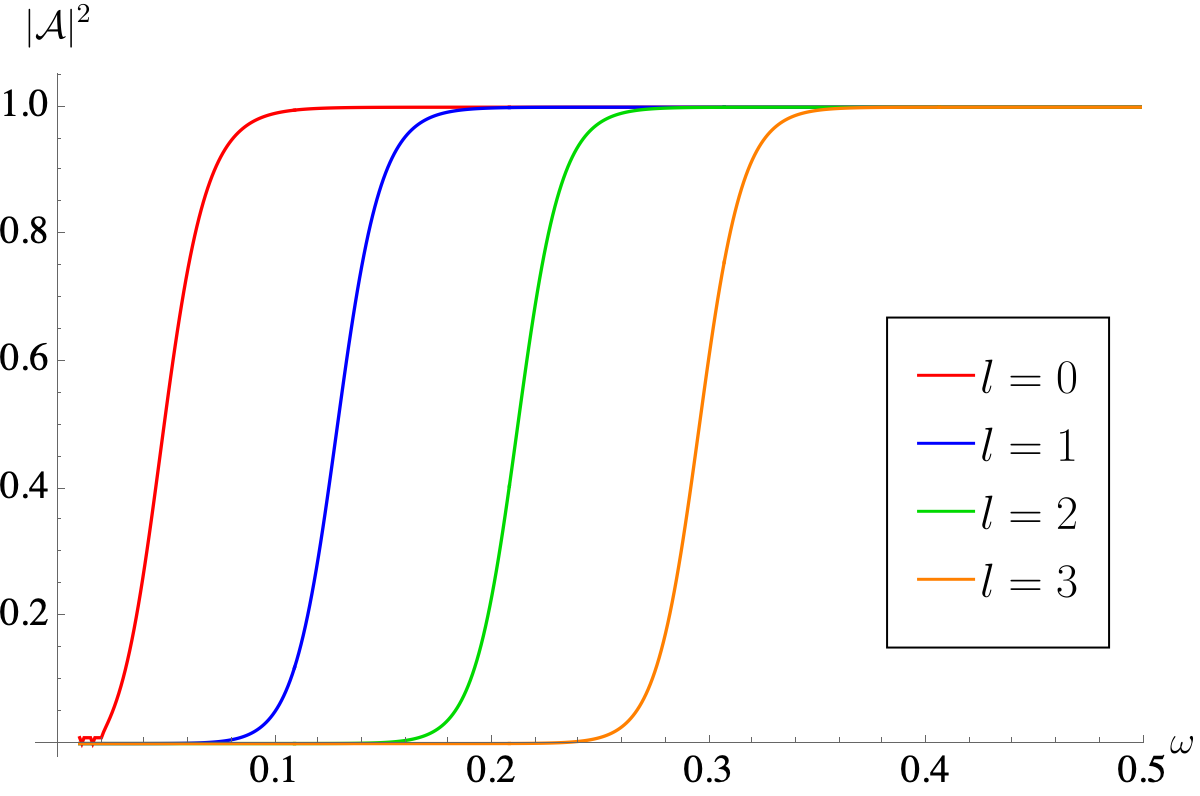}
    \end{subfigure}
    \hspace{0.5cm}
    \begin{subfigure}{0.47\textwidth}
    \centering
    \includegraphics[width=\textwidth]{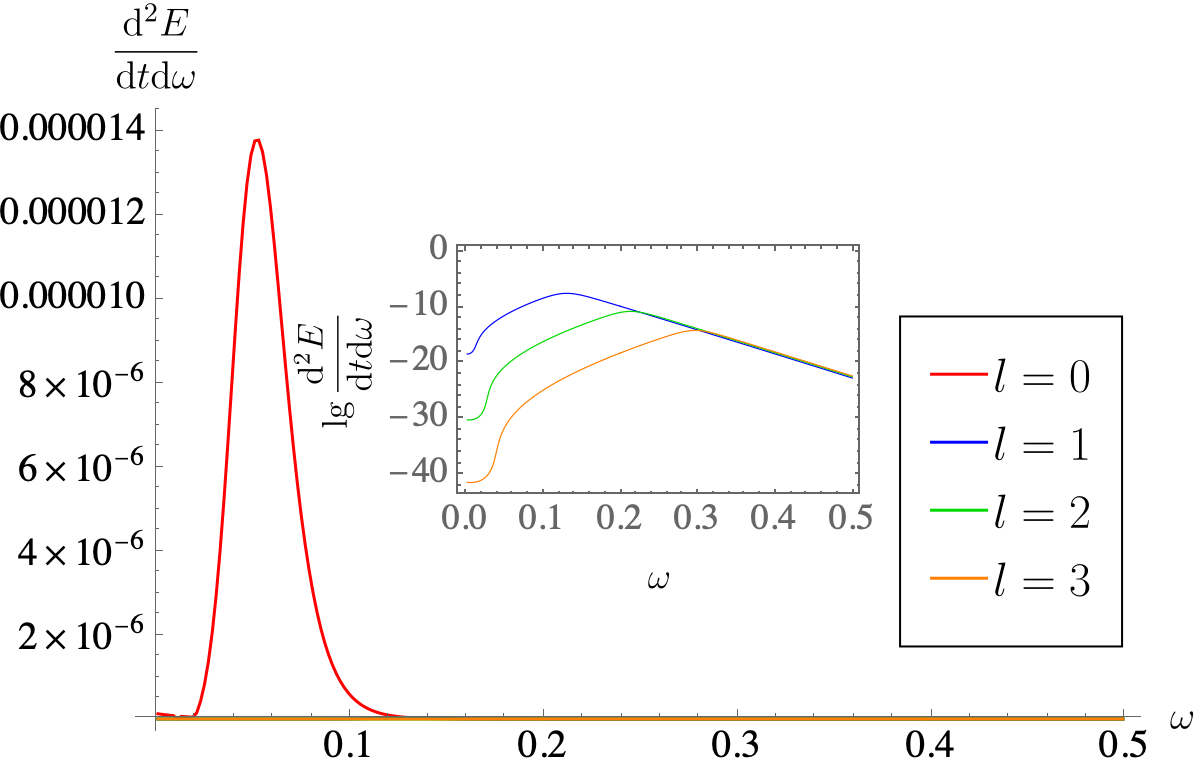}
    \end{subfigure}
    \caption{\raggedright Grey-body factor (left panel) and energy emission rate (right panel) as functions of $\omega$ for different $l$, with $\xi=5$ and $L=L_0$. The inset in the right panel shows the logarithmic plot of the emission rate to amplify the detailed features.}
    \label{fig:hawking_l_xi5}
\end{figure}

\begin{figure}[ht]
    \centering
    \begin{subfigure}{0.47\textwidth}
        \centering
        \includegraphics[width=\textwidth]{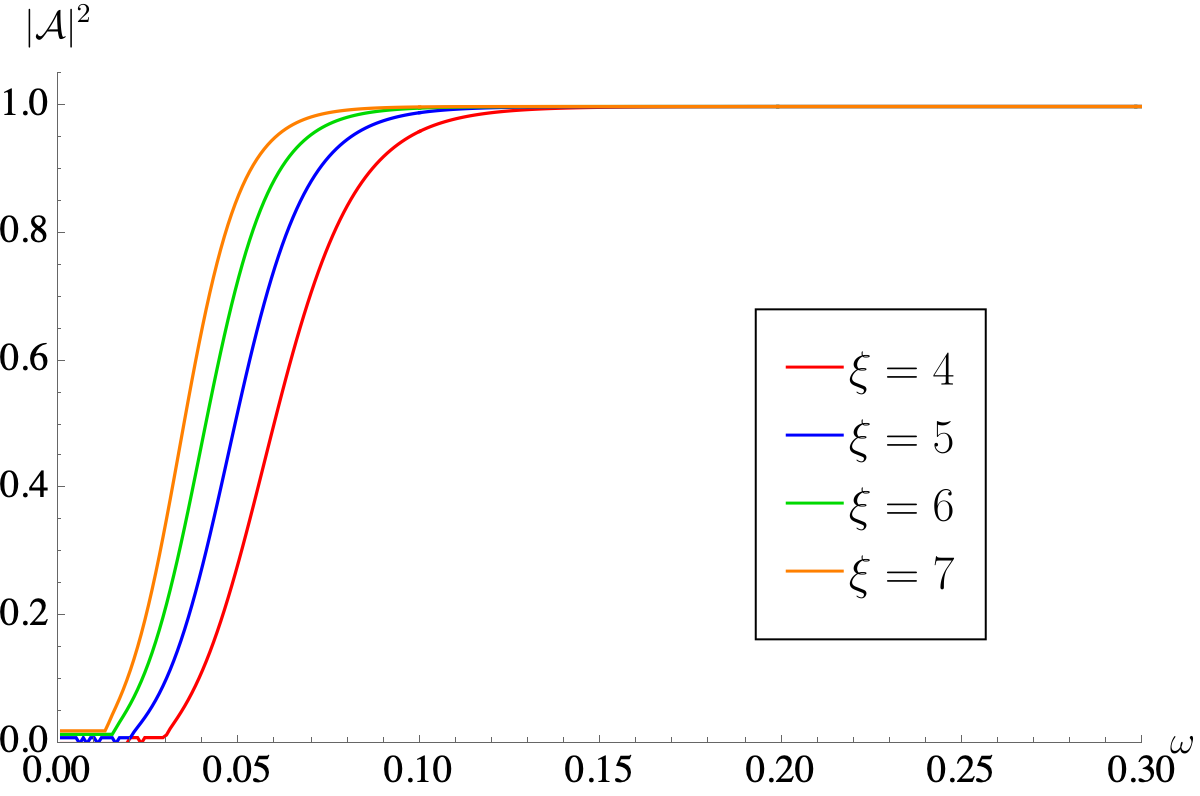}
    \end{subfigure}
    \hspace{0.5cm}
    \begin{subfigure}{0.47\textwidth}
    \centering
    \includegraphics[width=\textwidth]{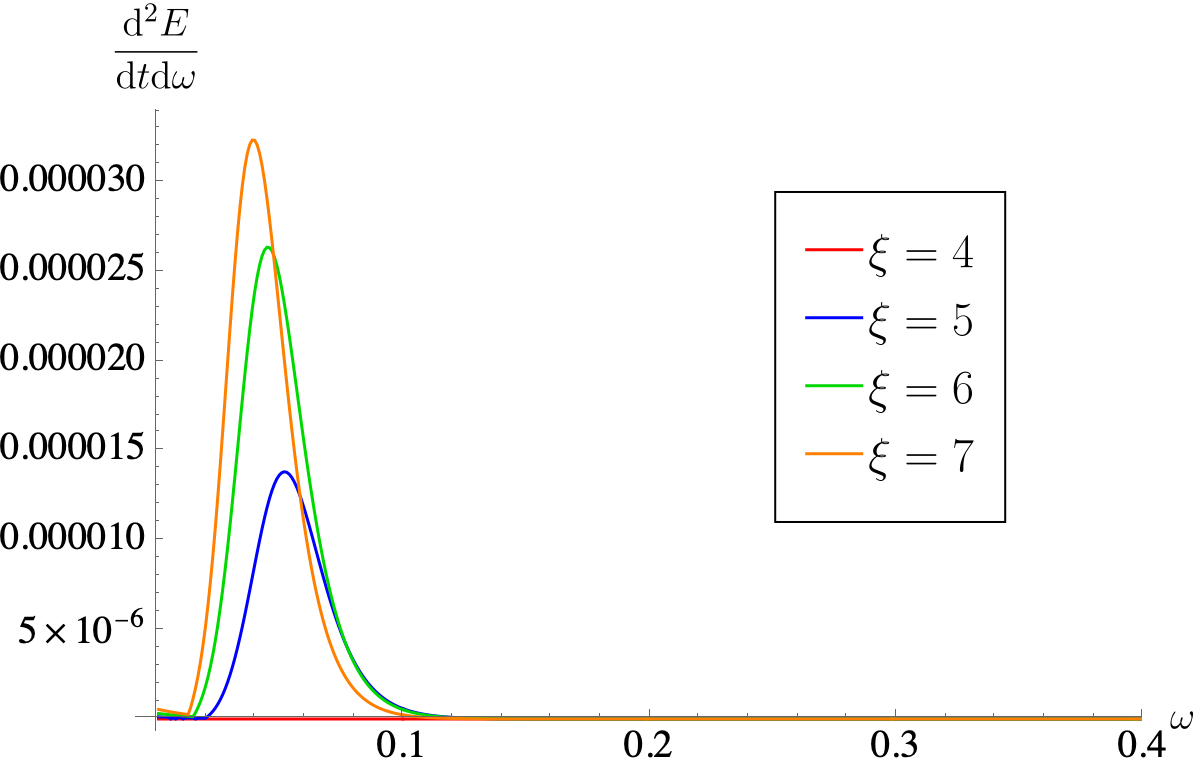}
    \end{subfigure}
    \caption{\raggedright Grey-body factor (left panel) and energy emission rate (right panel) as functions of $\omega$ for different $\xi$, with $l=0$ and $L=L_0$.}
    \label{fig:hawking_xi_l0}
\end{figure}

\begin{figure}[ht]
    \centering
    \begin{subfigure}{0.47\textwidth}
        \centering
        \includegraphics[width=\textwidth]{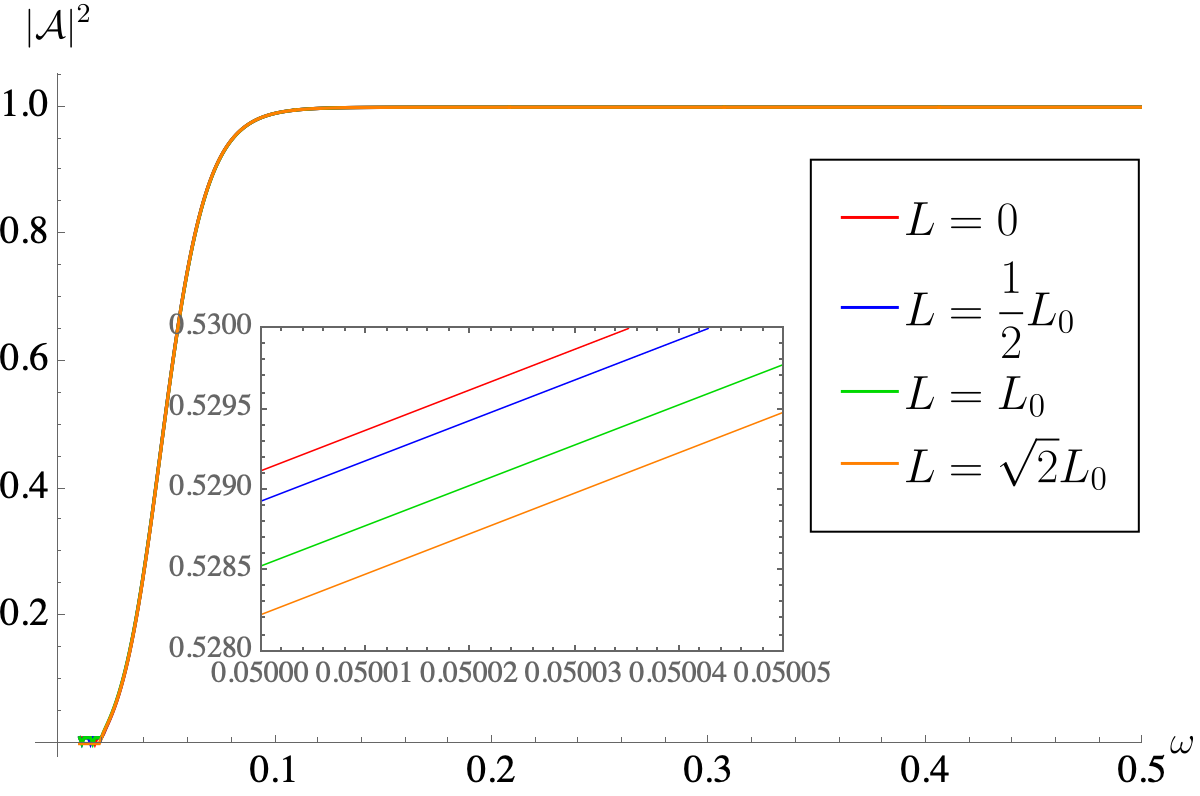}
    \end{subfigure}
    \hspace{0.5cm}
    \begin{subfigure}{0.47\textwidth}
    \centering
    \includegraphics[width=\textwidth]{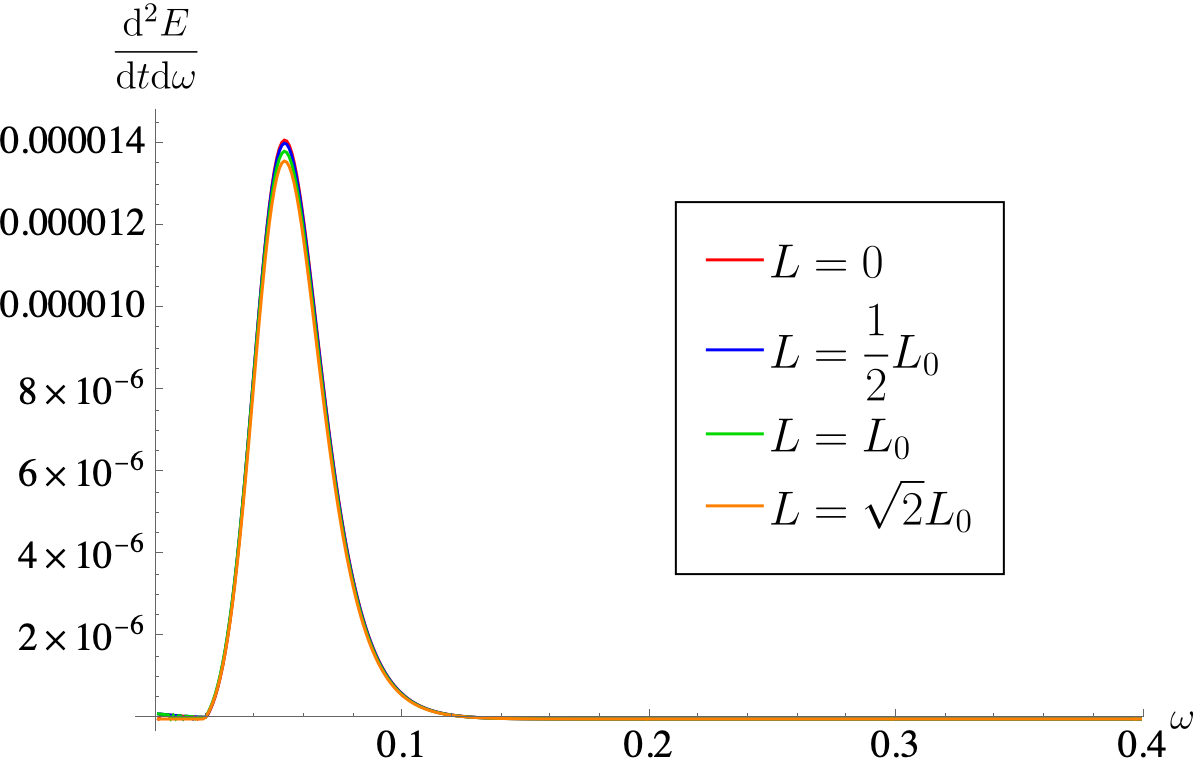}
    \end{subfigure}
    \caption{\raggedright Grey-body factor (left panel) and energy emission rate (right panel) as functions of $\omega$ for different $L$, with $l=0$ and $\xi=5$. A magnified view is shown in the inset of the left panel.}
    \label{fig:hawking_L_l0_xi5}
\end{figure}

The grey-body factors and the energy emission rates for the analogue Hawking radiation of the acoustic Hayward black hole are calculated using the 6th-order WKB approximation, as shown in Figs.~\ref{fig:hawking_l_xi5}, \ref{fig:hawking_xi_l0}, and \ref{fig:hawking_L_l0_xi5}. In these figures, the left panels show the grey-body factors, while the right panels display the energy emission rates, both plotted as functions of the frequency $\omega$ for different values of the angular momentum number $l$, the parameter $\xi$, and Hayward parameter $L$. Specifically, Fig.~\ref{fig:hawking_l_xi5} corresponds to fixed parameters $\xi=5$ and $L=L_0$, with results displayed for different values of $l=0,1,2,3$. Figs.~\ref{fig:hawking_xi_l0} show the behavior for $L=L_0$ and $l=1$, considering different values of $\xi=4,5,6,7$. Meanwhile, Fig.~\ref{fig:hawking_L_l0_xi5} displays the results for fixed $l=0$ and $\xi=5$, showing the dependence on $L$ with values $L=0,\frac{1}{2}L_0,L_0,$ and $\sqrt{2}L_0$. 

Based on the results presented, the following conclusions can be summarized:
\begin{itemize}
    \item In all cases, the grey-body factors for each mode begin at zero in the low frequency limit and asymptotically approach unity as the frequency increases. This trend reflects the fact that higher-frequency particles possess greater energy, making them more likely to tunnel through the potential barrier. Similarly, for every mode, the energy emission rate exhibits a distribution analogous to the blackbody radiation spectrum: it vanishes at both the low and high frequency limits, with a single peak occurring at intermediate frequencies. This behavior arises because the grey-body factor approaches zero at low frequencies, while the probability of producing high-energy particles becomes extremely small in the high-frequency regime.
    \item In Fig.~\ref{fig:hawking_l_xi5}, the left panel illustrates that as the angular momentum number $l$ increases, the grey-body factor exhibits a slower saturation. This behavior is explained by the rise in the effective potential barrier height with $l$, as shown in Fig.~\ref{fig:poten_l}, and indicates that the dominant contribution to the analogue Hawking radiation originates from the zero angular momentum modes.
    \item In Fig.~\ref{fig:hawking_xi_l0}, for a fixed angular momentum number $l$, both the grey-body factor and the energy emission rate increase with $\xi$. This behavior arises because a larger value of $\xi$ corresponds to a lower effective potential barrier, as shown in Fig.~\ref{fig:poten_xi}.
    \item In Fig.~\ref{fig:hawking_L_l0_xi5}, we examine how the Hayward parameter $L$ influences the Hawking radiation. It is shown that variations in $L$ induce only minimal changes in both the grey-body factors and the energy emission rates. This behavior is due to the fact that $L$ is constrained to the narrow range $0\leq L\leq\sqrt{2}L_0$, which limits its influence on the acoustic Hayward black hole configuration. This observation aligns with the conclusions drawn from Figs.~\ref{fig:horizon}, \ref{fig:sha_pic}, and \ref{fig:haw_temp}.
\end{itemize}

\section{Conclusions and discussions}\label{sec:con}
In this paper, we constructed the acoustic black hole within the Hayward spacetime using the relativistic GP theory, which presented the first extension of acoustic black holes to a regular black hole spacetime. We analyzed the horizon structure of this acoustic Hayward black hole, and investigated how the acoustic and Hayward horizons vary with the tuning parameter $\xi$ and the Hayward parameter $L$. It was shown that the acoustic horizon size increases with $\xi$ while slightly decreasing with increasing $L$. Based on this, we investigated the acoustic shadow of the acoustic Hayward black hole, the formation mechanism of the black hole shadow was explained and its radius was computed by analyzing the critical null geodesics. Our results showed that shadow images are sketched for different values of $\xi$ and $L$, which indicated that the shadow size increases with $\xi$ but slightly decreases with larger $L$.

Moreover, we analyzed the QNMs and analogue Hawking radiation of the acoustic Hayward black hole. By reducing the covariant scalar field equation near the black hole to a Schr\"{o}dinger-like form, the effective potential was shown to exhibit a potential barrier like profile for all parameter values considered, vanishing both at infinity and at the horizon. In addition, the height of the effective potential decreases as $\xi$ increases, rises with the angular momentum number $l$, and shows only weak dependence on the Hayward parameter $L$. These features are important for the subsequent calculation of QNMs and analogue Hawking radiation.

Then by imposing appropriate boundary conditions on the scalar field equation, we computed the QNM frequencies numerically using the WKB method. A WKB expansion of the scalar field yields the scattering matrix, and applying the boundary conditions on the scattering matrix leads to a WKB expression for the QNM frequency $\omega$. Implementing the 9th-order WKB approximation in Mathematica, we obtained the QNMs for different overtone numbers $n$ and angular momentum numbers $l$. These results are further validated with the AIM. It was shown that the real part of the QNM frequency is positive and the imaginary part negative for all modes considered, which confirms the stability of the acoustic Hayward black hole. As $l$ increases and $n$ decreases, the real part of the QNM frequency grows, while the magnitude of its imaginary part diminishes. This trend demonstrates how the acoustic Hayward black hole affects scalar fields of different modes, consistent with the corresponding variations in the effective potential.

Finally, by applying scattering boundary conditions to the phonon, we investigate the analogue Hawking radiation of the acoustic Hayward black hole. We examined how the analogue Hawking temperature depends on the parameter $\xi$ and $L$, and showed that the temperature initially increases and then decreases with rising $\xi$, while it exhibits a gradual decline as $L$ increases. Using the 6th-order WKB approximation, we computed the grey-body factors and energy emission rates. It was shown that an increase in the angular momentum number $l$ leads to a suppression of both the grey-body factor and the emission rate, which is explained by the corresponding change in the effective potential. Notably, the dominant contribution to the analogue Hawking radiation comes from the $l=0$ mode. As the parameter $\xi$ increases, the grey-body factors and emission rates are enhanced, a trend that follows the rise in temperature and is likewise linked to the evolution of the effective potential. In contrast, the parameter $L$ has a minimal effect on the analogue Hawking radiation due to its constrained range of variation. 

Our results demonstrated that the acoustic shadow provides a directly observable quantity, offering a potential experimental probe for simulating and detecting acoustic black holes. By analyzing the acoustic shadow, one can effectively uncover the geometric and dynamical properties near the acoustic horizon, thereby deepening the understanding of gravitational effects in the vicinity of black holes. This work not only opened new pathways for studying black hole like phenomena in laboratory settings but also provided a unique framework for exploring sound wave propagation in complex media. It facilitated a detailed comparison between acoustic fluid systems and the spacetime geometry of black holes in general relativity. 

Building on the current findings, several promising directions for future research emerge. First, it would be highly valuable to develop observational diagnostics capable of distinguishing between regular and singular acoustic black holes. A systematic comparison of acoustic Hayward black holes and acoustic Schwarzschild black holes should be conducted, with emphasis on their QNMs and shadow structures. Second, extending the acoustic regular black hole into the holographic framework and studying its connection with real gravity represents a natural and promising direction~\cite{Ge:2015uaa,Yu:2017bnu}. Third, constructing models of rotating acoustic regular black holes, moving beyond static and spherically symmetric configurations, represents an important generalization of the present work. Finally, a detailed investigation of the correlation between QNMs and the shadow radius of acoustic Hayward black hole is warranted~\cite{Stefanov:2010xz}. In the near future, it is anticipated that our theoretical results can provide potential applications in observations of astrophysical black holes.
\section*{Acknowledgement}
This work was supported by the National Natural Science Foundation of China (No.~12475069) and Guangdong Basic and Applied Basic Research Foundation (No.~2025A1515011321).





\end{document}